\documentclass[aps,prl,amsmath,amssymb,floatfix,twocolumn,superscriptaddress]{revtex4-1}
\usepackage{parskip} 
\usepackage{tikz} 
\usepackage{graphicx,ulem}
\usepackage{subfigure}
\usepackage{svg}

\usepackage{amsmath}
\usepackage{bbold}
\usepackage{slashed}
\usepackage{amssymb}
\usepackage{indentfirst}
\usepackage{braket}
\usepackage{array}
\usepackage[colorlinks]{hyperref}
\usepackage{float}
\usepackage[margin=1in]{geometry}

\usepackage{natbib}
\usepackage{xcolor}
\usepackage{hyperref}
\definecolor{Color}{RGB}{0, 191, 255} 
\hypersetup{colorlinks=true, citecolor=Color}

\begin{document}
	
	\preprint{APS/123-QED}
	\title{Topological Anderson Phase Transitions in Y-shaped Plasmonic Valley Metal-slabs}
	
	\author{Huichang Li}
	\affiliation{School of Physics and Materials Science, Nanchang University, Nanchang 330031, China}
	
	\author{Yun Shen}\thanks{shenyun@ncu.edu.cn}
	\affiliation{School of Physics and Materials Science, Nanchang University, Nanchang 330031, China}
	\affiliation{Institute of Space Science and Technology, Nanchang University, Nanchang 330031, China}
	
	\author{Xiaohua Deng}\thanks{dengxhua@gmail.com}
	\affiliation{Institute of Space Science and Technology, Nanchang University, Nanchang 330031, China}
	
	\date{\today}
	
	\begin{abstract} 
		
	Throughout history, all developmental trajectories of civilization - encompassing progress, creation, and innovation - have fundamentally pursued the paradigm shift 'from disorder to order'. In photonics, investigations into disordered systems have primarily focused on foundational principles governing signal diffusion and localization. This paper addresses terahertz device development by examining the dual role of disorder in photonic systems: while potentially compromising optical transmission stability, it simultaneously inspires innovative topological protection mechanisms. Building upon the symmetry-breaking induced valley-Hall topological Anderson phase transition in Y-shaped metallic structures, we achieve valley Chern number modulation through random rotation of constituent units, demonstrating progressive emergence of in-gap topological states with increasing disorder parameters and observing topological negative refraction phenomena. Furthermore, an effective Dirac two-band model is established to quantitatively characterize the evolution of bulk transport states under disorder variation. By strategically regulating disordered configurations to induce valley-Hall topological Anderson phase transitions, this research provides new pathways for overcoming critical technical challenges in terahertz devices, particularly transmission loss limitations.
	
	\end{abstract}
	
	\maketitle
	
	\par 
	{\bf Introduction}: The study of disorder phenomena in photonics is of great significance for understanding light transmission mechanisms and device development~\cite{1,2,3,4,5}, such as spatial light modulators~\cite{6,7,8}, random lasers~\cite{9,10,11,12}, and photonic crystal waveguides~\cite{13,14,15,16}. Although disordered structures often cause negative effects like backscattering in optical communications, the unique phenomena they induce (e.g., Anderson localization) reveal entirely new physical laws governing light-matter interactions~\cite{17,18,19,20,21,22}. Notably, topological photonics, by constructing protected edge states, enables stable light transmission even in disordered environments—a property that demonstrates unique value in terahertz frequency research~\cite{23,24,25,26,27,28,29,30,31,32,33,34,35,36,37,38}. The defining characteristic of a topological Anderson insulator is the localization of wavefunctions in disordered systems, and its hallmark is the disorder-induced transition of a system from a topologically trivial to a topologically nontrivial state~\cite{17,21,39,40,41,42,43,44,45,46}.
	
	\par 
	Terahertz devices face challenges such as high transmission losses and difficulties in miniaturization~\cite{47,48,49,50,51,52,53,54,55,56}, while the controlled regulation of disorder distributions offers a new approach to address these issues. By designing specific disordered structures, precise control of terahertz waveguides can be achieved; however, it is essential to simultaneously investigate the degradation mechanisms of disorder on device performance and develop suppression techniques. In terms of applications, the synergistic effects of spoof surface plasmon polaritons (SSPPs) with disorder have garnered significant attention~\cite{57,58,59,60,61,62,63,64}. By introducing controllable disorder, the subwavelength localization properties of SSPPs can be significantly enhanced, opening new pathways for enhancing light-matter interactions in the terahertz band. 
	
	\par 
	In this paper, the topological Anderson phase transition originates from the broken symmetry of the Y-shaped metallic SSPPs structure, leading to a valley-Hall topology. Specifically, we constructed a domain wall waveguide and introduced geometric disorder by randomly rotating all individual unit cells on one side of the domain wall. As the disorder strength increases, this side transitions from a topologically trivial phase to a topologically nontrivial phase. The hallmark of this topological Anderson phase transition is the transition from a zero to a non-zero valley Chern number. Additionally, we demonstrated topological negative refraction under topologically nontrivial disorder conditions and quantitatively characterized the disorder-induced bulk state evolution of the domain wall waveguide by constructing an effective model, obtaining the expected phase transition point.
	
	\par
\begin{figure*}[htb]
		\centering
		\includegraphics[width=9cm]{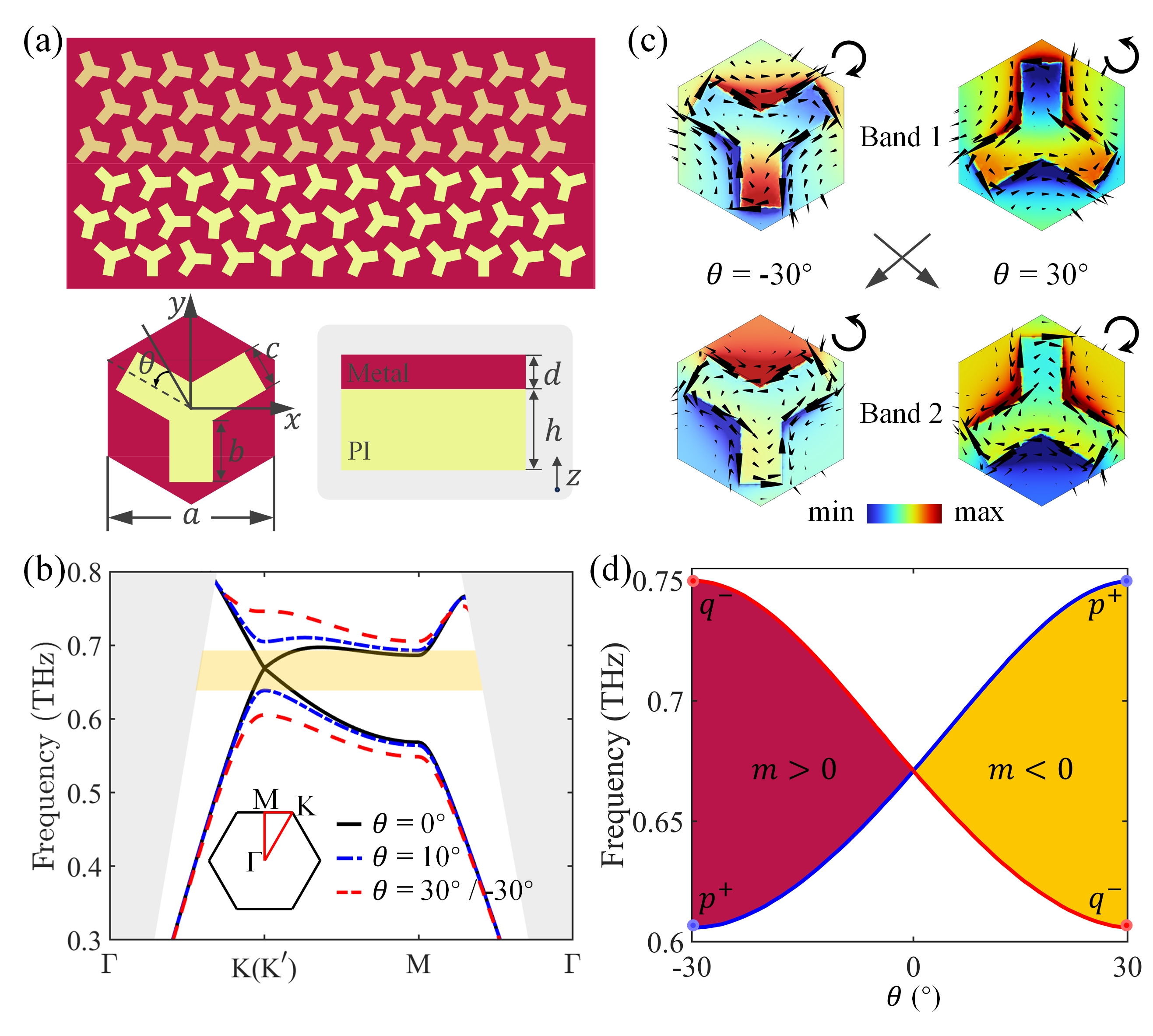}
		\caption{(a) Schematic diagram of the YTAPC waveguide system, unit cell and cross-sectional layered structure along the $z$-direction. (b) The black solid line, blue dotted line, and red dashed line represent the band diagrams of the unit cell for $\theta$=$0^{\circ}$, $\theta$=$10^{\circ}$, $\theta$=$30^{\circ}$/$-30^{\circ}$, respectively. The yellow region indicates the bandgap that appears at $\theta$=$10^{\circ}$. (c) The evolution of the $E_{z}$ field eigenstates from $\theta$=$-30^{\circ}$ to $\theta$=$30^{\circ}$, and (d) the illustration of band inversion and topological phase transition at the $K/K'$ points as a function of $\theta$.		
		}
		\label{fig1}
	\end{figure*}

 
	{\bf Design and theory}: As shown in fig.~\ref{fig1}(a), the Y-shaped-metal SSPPs-based Topological Anderson Photonic Crystal (YTAPC) waveguide system. The unit cell of the photonic lattice consists of three identical rectangular air holes, with a length of b = 60 \textmu m, a width of c = 35 \textmu m, and a lattice constant of a = 135 \textmu m. The thickness of the metal and the substrate Polyimide (PI) are d = 1 \textmu m and h = 50 \textmu m, respectively. In our simulation, the dielectric constant of PI is set to $\epsilon = 3.5$, and the metal is set to PEC (Perfect Electric Conductor) conditions.

	\par 
	For the YTAPC photonic crystal lattice unit cell with air hole rotation angles $\theta$ of $-30^{\circ}$, $0^{\circ}$, and $30^{\circ}$, the photonic bands for TE modes (electric field along the $z$-axis) are shown in fig.~\ref{fig1}(b). Results indicate that at $\theta$=$0^{\circ}$, the bandgap closes at the valley $K/K'$ points due to $C_{3v}$ symmetry, whereas at $\theta$=$-30^{\circ}$ and $\theta$=$30^{\circ}$, the bandgap opens, indicating symmetry breaking and a topological phase transition. Figure.~\ref{fig1}(c) illustrates the variation of the $E_{z}$ field at the first and second characteristic frequencies at the $K$ point for $\theta$=$-30^{\circ}$ and $\theta$=$30^{\circ}$. The left or right circular polarization power flux of the eigenstates clearly indicates the occurrence of band inversion and a topological phase transition.

	\par 
	{\bf SSPPs-based Valley-Hall Topology}: Before entering the analysis of topological Anderson phase transitions, it is essential to first clarify the topological transport of YTAPC.
	
	\par 
	The topological properties of photonic structures based on valley-Hall were elaborated in the previous section, characterized by the valley Chern number. As depicted in fig.~\ref{fig1}(d), during $\theta$ from $-30^{\circ}$ to $30^{\circ}$, a topological phase transition occurs between the two lowest characteristic frequencies. The sign of the mass term $m$ in the equivalent Hamiltonian of the system changes before and after the phase transition, leading to a change in the sign of the valley Chern number, from $1/2$ before the transition to $-1/2$ after the transition.
	
	\par
\begin{figure*}[htb]
		\centering
		\includegraphics{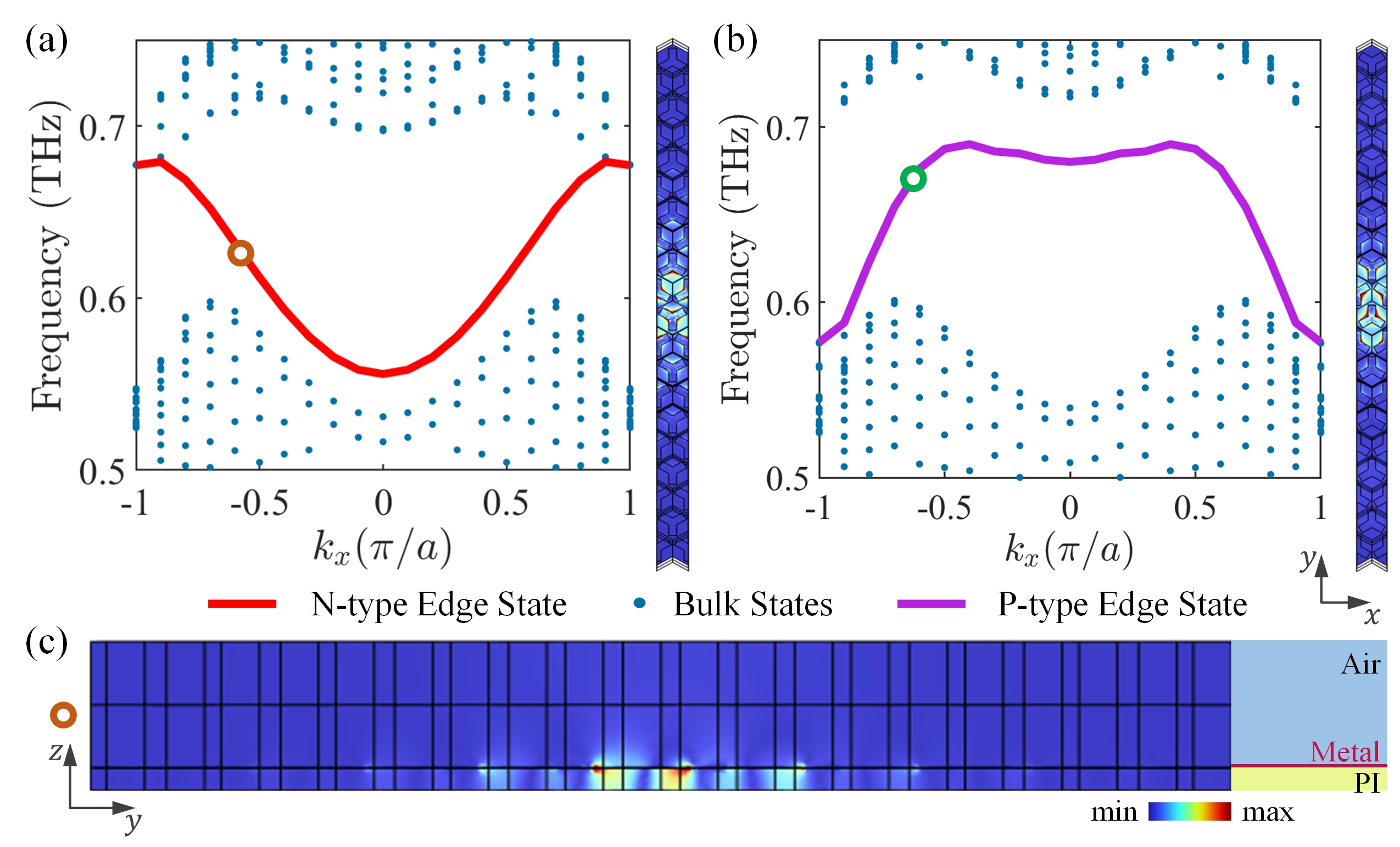}
		\caption{(a) and (b) show the supercell band diagrams for N-type and P-type configurations, respectively, with the marked positions on the right indicating the $|E_{z}|$ field distributions of the eigenstates. Cyan dots, red solid lines, and purple solid lines denote the bulk states, N-type topological edge states, and P-type topological edge states, respectively. (c) the $|E_{z}|$ field distribution in the $y$-$z$ plane corresponding to the marked points in (a).}
		\label{fig2}
	\end{figure*}

 
	It is well-known that when the difference in valley Chern numbers between two photonic structures constituting a domain wall is $\Delta C=1$, the domain wall supports valley-Hall topological edge states. The topological edge states within the bandgap supported by the domain wall between YTAPCs with $\theta$ of $-30^{\circ}$ and $30^{\circ}$ are shown in the supercell photonic bands of fig.~\ref{fig2}(a) and (b), where the upper/lower: $-30^{\circ}$/$30^{\circ}$ (N-type) and $30^{\circ}$/$-30^{\circ}$ (P-type) have frequency bands occupied by topological states from 0.6 THz to 0.68 THz and 0.6 THz to 0.69 THz, respectively. It is important to note that SSPPs exhibit remarkable localization properties. Figure.~\ref{fig2}(c) illustrates the highly localized $|E_{z}|$ field strength of the eigenstate corresponding to the circled excited topological state in fig.~\ref{fig2}(a), at the interface between the 1 \textmu m thick metal and PI.

	\par
\begin{figure*}[htb]
		\centering
		\includegraphics{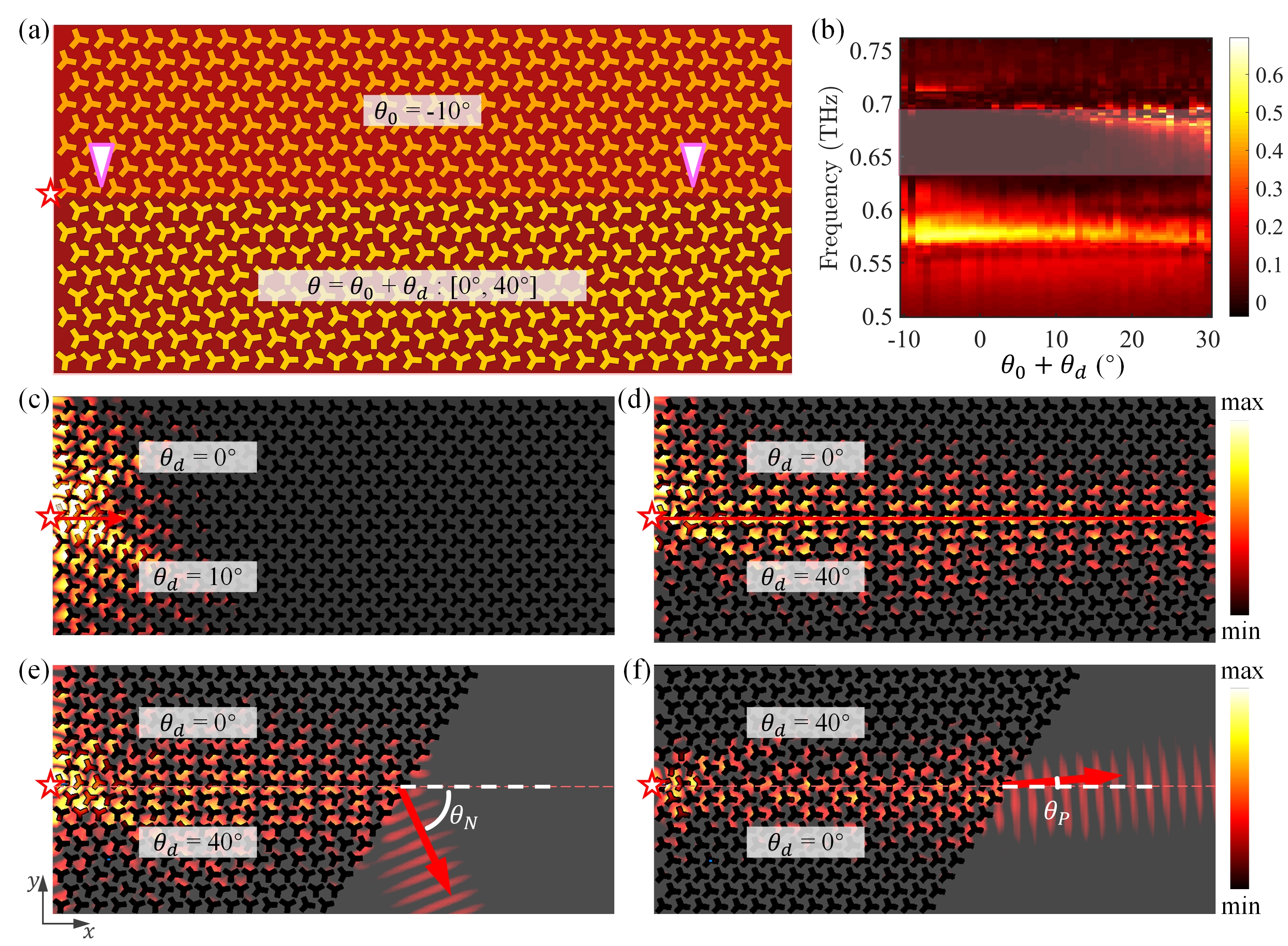}
		\caption{Topological Anderson phase transition induced by disorder. (a) Schematic of the YTAPC domain wall waveguide, where the upper side of the domain wall has units rotated at $\theta_{0}$=$-10^{\circ}$, and the lower side units have rotation angles of $\theta=\theta_{0}+\theta_{d}$, with $\theta_{d}: [0^{\circ}, 40^{\circ}]$ representing the disorder factor. (b) Transmittance phase diagram of the waveguide domain wall system as a function of $\theta_{d}$. (c) Upper/lower: $|E_{z}|$ field distribution at $\theta_{d}$=$0^{\circ}$/$\theta_{d}$=$10^{\circ}$ and (d) Upper/lower: $|E_{z}|$ field distribution at $\theta_{d}$=$0^{\circ}$/$\theta_{d}$=$40^{\circ}$, illustrating the transmission modes at 0.66 THz; (e) Refraction characteristics of the topological edge modes for N-type, upper/lower: $\theta_{d}$=$0^{\circ}$/$\theta_{d}$=$40^{\circ}$, and (f) for P-type, upper/lower: $\theta_{d}$=$40^{\circ}$/$\theta_{d}$=$0^{\circ}$, showing the $|E_{z}|$ field distribution at 0.66 THz.}
		\label{fig3}
	\end{figure*}

	{\bf Topological Anderson Phase Transitions Induced by Disorder}: Next, we explore the topological Anderson phase transitions induced by disorder. As shown in fig.~\ref{fig3}(a), we designed a domain wall waveguide system where the upper side of the domain wall is composed entirely of unit structures with $\theta_{0}$=$-10^{\circ}$, while the lower side features units with rotation angles based on $\theta_{0}$ plus a random factor $\theta_{d}$, chosen within the range $[0^{\circ}, 40^{\circ}]$. When terahertz signals are incident from the left port of the domain wall, measuring the transmittance at two points yields the transmittance phase diagram shown in fig.~\ref{fig3}(b) as $\theta_{d}$ varies. Notably, for small $\theta_{d}$ (approximately $[0^{\circ}, 20^{\circ}]$), the gray bandgap represents an ordinary insulating state. However, as $\theta_{d}$ increases to about $[20^{\circ}, 40^{\circ}]$, conductive transmission states emerge within the gray bandgap, and their frequency range broadens with increasing $\theta_{d}$. Further analysis at $\theta_{d}$=$10^{\circ}$ and $\theta_{d}$=$40^{\circ}$ for 0.66 THz reveals $|E_{z}|$ field distributions in fig.~\ref{fig3}(c) and (d). The former shows no transmission, indicating a topologically trivial bandgap, while the latter shows transmission localized within the domain wall, signifying disorder-induced valley-Hall topological Anderson transport. This confirms that the domain wall waveguide transitions from a topologically trivial to a nontrivial state as the disorder factor increases.

	\par 
	We also explored the valley topological refraction of YTAPC for potential applications in photonic on-chip system. As shown in fig.~\ref{fig3}(e) and (f), we constructed domain wall waveguide structures based on the N-type and P-type configurations illustrated in fig~\ref{fig2}. At the output interface of the domain wall (zigzag-type boundary), the incident wavevector $k$ can match the isofrequency curve of the background material via $k\cdot e^{t} = K\cdot e^{t}$, determining the direction of the outgoing light beam. Specifically, using $|k|\cdot cos(120^{\circ} + \theta_{N}) = |K|\cdot cos(60^{\circ})$, $|k|\cdot cos(60^{\circ} - \theta_{P}) = |K|\cdot cos(60^{\circ})$, and the wavevector forms $|k|= \frac{2 \pi \cdot (f \cdot n_{PI})}{c}$ and $|K|= \frac{2}{3} \cdot \frac{2 \pi}{a}$, we calculated the theoretical refraction angles as $\theta_{N} \approx -70^{\circ}$ and $\theta_{P} \approx 6.9^{\circ}$, corresponding to negative and positive topological transmission modes, respectively.
	
	\par
\begin{figure*}[htb]
		\centering
		\includegraphics{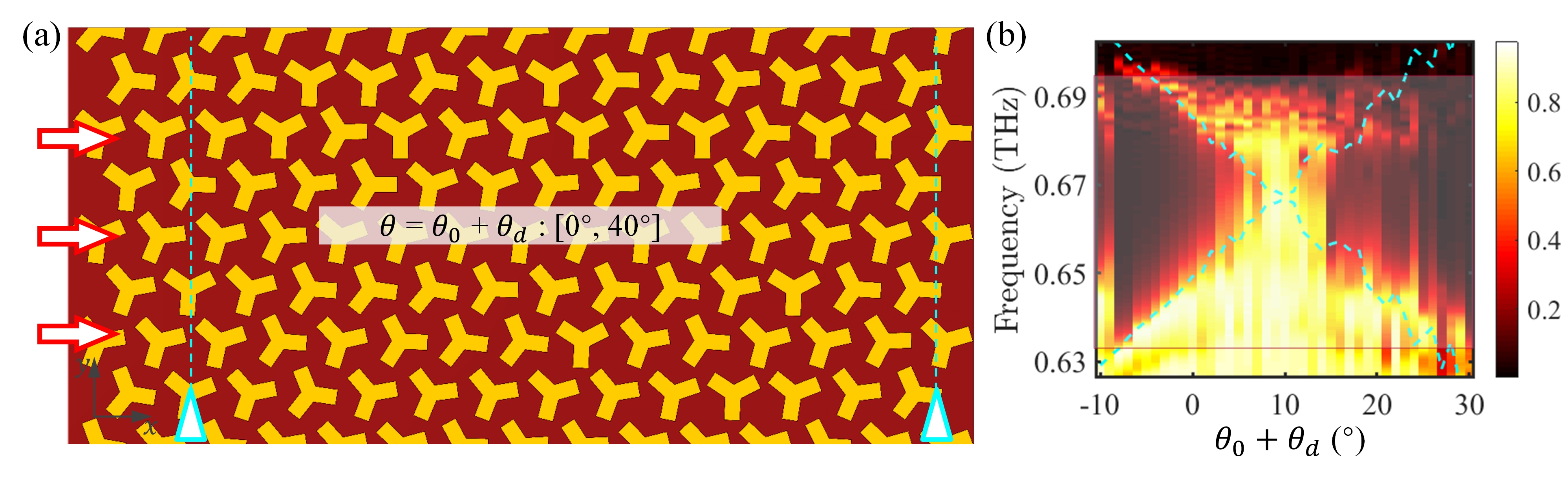}
		\caption{(a) Schematic of the pure disordered topological waveguide rotation angles of with unit cell: $\theta=\theta_{0}+\theta_{d}$ ($\theta_{d}: [0^{\circ}, 40^{\circ}]$), where cyan dashed lines mark the monitor positions; (b) Transmittance phase diagram of the waveguide system in (a) as $\theta_{d}$ varies, with cyan dashed lines indicating the characteristic frequency evolution calculated using the Hamiltonian effective model in eq.~\ref{eq1}.}
		\label{fig4}
	\end{figure*}

 
	Figure.~\ref{fig4}(a) shows a pure disordered waveguide system without domain wall, designed to further investigate the bulk state transport properties of the disorder factor. Unlike fig.~\ref{fig3}(a), this system has periodic boundary conditions in the $y$-direction. Numerical simulations yielded the transmittance phase diagram shown in fig.~\ref{fig4}(b), where the bandgap closes near $\theta_{d} = 20^{\circ}$ as $\theta_{d}$ increases from $0^{\circ}$ to $40^{\circ}$.
	
	\par 
	To quantitatively analyze this disorder-induced phenomenon, we constructed an effective Dirac two-band model:	
	\begin{equation}
		\label{eq1}
		H_d = \begin{pmatrix}
			f_0 + m_0 + \frac{\sum_n m_{d,n}}{N} & t + t \cdot e^{-ik} \\
			t + t \cdot e^{ik} & f_0 - m_0 - \frac{\sum_n m_{d,n}}{N}
		\end{pmatrix},
	\end{equation}
	where $f_{0} \approx 0.6719$ THz and $m_{0} \approx 0.039$ are fitting constants, $t$ represents the coupling strength, $N = 116$ is the number of unit cells in the disordered system, and $m_{d,n}$ denotes the disorder rotation angle introduced in the $n$-th unit cell. By gradually increasing $m_{d}$ and substituting values into this Hamiltonian model, we obtained the phase transition diagram represented by the cyan dashed line in fig.~\ref{fig4}(b), which aligns well with the simulation results, confirming the model's predictive accuracy for the evolution of bulk states and bandgaps.
	
	\par
	Using this effective Hamiltonian model, we found that the impact of the disorder factor is equivalent to the average of the disorder factors across all unit cells in the waveguide system. Thus, the transmission closure point (i.e., the bandgap closure point) in fig.~\ref{fig4}(b) is located at approximately $\theta = -10^{\circ} + \frac{\sum_{n} \theta_{d,n}}{N} = 0^{\circ}$, leading to $\theta_{d} = 20^{\circ}$.
		
	\par
	{\bf Conclusion}: In summary, this paper investigates valley-Hall topological Anderson phase transitions induced by rotational disorder factors. By constructing a domain wall waveguide system and introducing geometric disorder through the rotation angles of unit cells on one side of the domain wall, we observe a change in the overall Chern number sign as the disorder strength increases, leading to a transition from a topologically trivial to a nontrivial phase. Additionally, we demonstrate topological negative refraction in the nontrivial phase and develop a Hamiltonian-based two-band model to quantitatively describe the evolution of photonic transport states induced by disorder, accurately predicting the topological phase transition points. These disorder-induced topological edge states offer a viable approach for manipulating on-chip Terahertz wave propagation and hold promise for the development of efficient transmission devices. Notably, compared to existing topological photonic chip designs, our SSPPs configuration further reduces device thickness, facilitating integration and supporting the advancement of high-speed wireless communication from 6G to XG.

{\bf Acknowledgments}: This work was supported by the National Natural Science Foundation of China (Grant numbers 61927813, 62165008).

{\bf Data availability statement}:
The data that support the findings of this study are available upon reasonable request from the authors.

{\bf Conflict of Interest}:
The authors declare no competing financial interests.

\bibliography{Reference.bib}

\begin{thebibliography}{64}%
\makeatletter
\providecommand \@ifxundefined [1]{%
 \@ifx{#1\undefined}
}%
\providecommand \@ifnum [1]{%
 \ifnum #1\expandafter \@firstoftwo
 \else \expandafter \@secondoftwo
 \fi
}%
\providecommand \@ifx [1]{%
 \ifx #1\expandafter \@firstoftwo
 \else \expandafter \@secondoftwo
 \fi
}%
\providecommand \natexlab [1]{#1}%
\providecommand \enquote  [1]{``#1''}%
\providecommand \bibnamefont  [1]{#1}%
\providecommand \bibfnamefont [1]{#1}%
\providecommand \citenamefont [1]{#1}%
\providecommand \href@noop [0]{\@secondoftwo}%
\providecommand \href [0]{\begingroup \@sanitize@url \@href}%
\providecommand \@href[1]{\@@startlink{#1}\@@href}%
\providecommand \@@href[1]{\endgroup#1\@@endlink}%
\providecommand \@sanitize@url [0]{\catcode `\\12\catcode `\$12\catcode
  `\&12\catcode `\#12\catcode `\^12\catcode `\_12\catcode `\%12\relax}%
\providecommand \@@startlink[1]{}%
\providecommand \@@endlink[0]{}%
\providecommand \url  [0]{\begingroup\@sanitize@url \@url }%
\providecommand \@url [1]{\endgroup\@href {#1}{\urlprefix }}%
\providecommand \urlprefix  [0]{URL }%
\providecommand \Eprint [0]{\href }%
\providecommand \doibase [0]{http://dx.doi.org/}%
\providecommand \selectlanguage [0]{\@gobble}%
\providecommand \bibinfo  [0]{\@secondoftwo}%
\providecommand \bibfield  [0]{\@secondoftwo}%
\providecommand \translation [1]{[#1]}%
\providecommand \BibitemOpen [0]{}%
\providecommand \bibitemStop [0]{}%
\providecommand \bibitemNoStop [0]{.\EOS\space}%
\providecommand \EOS [0]{\spacefactor3000\relax}%
\providecommand \BibitemShut  [1]{\csname bibitem#1\endcsname}%
\let\auto@bib@innerbib\@empty
\bibitem [{\citenamefont {Choi}\ \emph {et~al.}(2017)\citenamefont {Choi},
  \citenamefont {Jo}, \citenamefont {Ahn}, \citenamefont {Seo}, \citenamefont
  {Park}, \citenamefont {Jhon},\ and\ \citenamefont {Choi}}]{1}%
  \BibitemOpen
  \bibfield  {author} {\bibinfo {author} {\bibfnamefont {W.}~\bibnamefont
  {Choi}}, \bibinfo {author} {\bibfnamefont {Y.}~\bibnamefont {Jo}}, \bibinfo
  {author} {\bibfnamefont {J.}~\bibnamefont {Ahn}}, \bibinfo {author}
  {\bibfnamefont {E.}~\bibnamefont {Seo}}, \bibinfo {author} {\bibfnamefont
  {Q.~H.}\ \bibnamefont {Park}}, \bibinfo {author} {\bibfnamefont {Y.~M.}\
  \bibnamefont {Jhon}}, \ and\ \bibinfo {author} {\bibfnamefont
  {W.}~\bibnamefont {Choi}},\ }\href {\doibase 10.1038/ncomms14636} {\bibfield
  {journal} {\bibinfo  {journal} {Nat. Commun.}\ }\textbf {\bibinfo {volume}
  {8}},\ \bibinfo {pages} {14636} (\bibinfo {year} {2017})}\BibitemShut
  {NoStop}%
\bibitem [{\citenamefont {Garc{\'i}a}\ and\ \citenamefont {Lodahl}(2017)}]{2}%
  \BibitemOpen
  \bibfield  {author} {\bibinfo {author} {\bibfnamefont {P.~D.}\ \bibnamefont
  {Garc{\'i}a}}\ and\ \bibinfo {author} {\bibfnamefont {P.}~\bibnamefont
  {Lodahl}},\ }\href {\doibase https://doi.org/10.1002/andp.201600351}
  {\bibfield  {journal} {\bibinfo  {journal} {Ann. Phys.}\ }\textbf {\bibinfo
  {volume} {529}},\ \bibinfo {pages} {1600351} (\bibinfo {year}
  {2017})}\BibitemShut {NoStop}%
\bibitem [{\citenamefont {Gigan}(2022)}]{3}%
  \BibitemOpen
  \bibfield  {author} {\bibinfo {author} {\bibfnamefont {S.}~\bibnamefont
  {Gigan}},\ }\href {\doibase 10.1038/s41567-022-01681-1} {\bibfield  {journal}
  {\bibinfo  {journal} {Nat. Phys.}\ }\textbf {\bibinfo {volume} {18}},\
  \bibinfo {pages} {980} (\bibinfo {year} {2022})}\BibitemShut {NoStop}%
\bibitem [{\citenamefont {Vynck}\ \emph {et~al.}(2023)\citenamefont {Vynck},
  \citenamefont {Pierrat}, \citenamefont {Carminati}, \citenamefont
  {Froufe-P{\'e}rez}, \citenamefont {Scheffold}, \citenamefont {Sapienza},
  \citenamefont {Vignolini},\ and\ \citenamefont {S{\'a}enz}}]{4}%
  \BibitemOpen
  \bibfield  {author} {\bibinfo {author} {\bibfnamefont {K.}~\bibnamefont
  {Vynck}}, \bibinfo {author} {\bibfnamefont {R.}~\bibnamefont {Pierrat}},
  \bibinfo {author} {\bibfnamefont {R.}~\bibnamefont {Carminati}}, \bibinfo
  {author} {\bibfnamefont {L.~S.}\ \bibnamefont {Froufe-P{\'e}rez}}, \bibinfo
  {author} {\bibfnamefont {F.}~\bibnamefont {Scheffold}}, \bibinfo {author}
  {\bibfnamefont {R.}~\bibnamefont {Sapienza}}, \bibinfo {author}
  {\bibfnamefont {S.}~\bibnamefont {Vignolini}}, \ and\ \bibinfo {author}
  {\bibfnamefont {J.~J.}\ \bibnamefont {S{\'a}enz}},\ }\href {\doibase
  10.1103/RevModPhys.95.045003} {\bibfield  {journal} {\bibinfo  {journal}
  {Rev. Mod. Phys.}\ }\textbf {\bibinfo {volume} {95}},\ \bibinfo {pages}
  {045003} (\bibinfo {year} {2023})}\BibitemShut {NoStop}%
\bibitem [{\citenamefont {Yu}\ \emph {et~al.}(2021)\citenamefont {Yu},
  \citenamefont {Qiu}, \citenamefont {Chong}, \citenamefont {Torquato},\ and\
  \citenamefont {Park}}]{5}%
  \BibitemOpen
  \bibfield  {author} {\bibinfo {author} {\bibfnamefont {S.}~\bibnamefont
  {Yu}}, \bibinfo {author} {\bibfnamefont {C.-W.}\ \bibnamefont {Qiu}},
  \bibinfo {author} {\bibfnamefont {Y.}~\bibnamefont {Chong}}, \bibinfo
  {author} {\bibfnamefont {S.}~\bibnamefont {Torquato}}, \ and\ \bibinfo
  {author} {\bibfnamefont {N.}~\bibnamefont {Park}},\ }\href {\doibase
  10.1038/s41578-020-00263-y} {\bibfield  {journal} {\bibinfo  {journal} {Nat.
  Rev. Mater.}\ }\textbf {\bibinfo {volume} {6}},\ \bibinfo {pages} {226}
  (\bibinfo {year} {2021})}\BibitemShut {NoStop}%
\bibitem [{\citenamefont {Hsu}\ \emph {et~al.}(2017)\citenamefont {Hsu},
  \citenamefont {Liew}, \citenamefont {Goetschy}, \citenamefont {Cao},\ and\
  \citenamefont {Douglas~Stone}}]{6}%
  \BibitemOpen
  \bibfield  {author} {\bibinfo {author} {\bibfnamefont {C.~W.}\ \bibnamefont
  {Hsu}}, \bibinfo {author} {\bibfnamefont {S.~F.}\ \bibnamefont {Liew}},
  \bibinfo {author} {\bibfnamefont {A.}~\bibnamefont {Goetschy}}, \bibinfo
  {author} {\bibfnamefont {H.}~\bibnamefont {Cao}}, \ and\ \bibinfo {author}
  {\bibfnamefont {A.}~\bibnamefont {Douglas~Stone}},\ }\href {\doibase
  10.1038/nphys4036} {\bibfield  {journal} {\bibinfo  {journal} {Nat. Phys.}\
  }\textbf {\bibinfo {volume} {13}},\ \bibinfo {pages} {497} (\bibinfo {year}
  {2017})}\BibitemShut {NoStop}%
\bibitem [{\citenamefont {Jang}\ \emph {et~al.}(2018)\citenamefont {Jang},
  \citenamefont {Horie}, \citenamefont {Shibukawa}, \citenamefont {Brake},
  \citenamefont {Liu}, \citenamefont {Kamali}, \citenamefont {Arbabi},
  \citenamefont {Ruan}, \citenamefont {Faraon},\ and\ \citenamefont
  {Yang}}]{7}%
  \BibitemOpen
  \bibfield  {author} {\bibinfo {author} {\bibfnamefont {M.}~\bibnamefont
  {Jang}}, \bibinfo {author} {\bibfnamefont {Y.}~\bibnamefont {Horie}},
  \bibinfo {author} {\bibfnamefont {A.}~\bibnamefont {Shibukawa}}, \bibinfo
  {author} {\bibfnamefont {J.}~\bibnamefont {Brake}}, \bibinfo {author}
  {\bibfnamefont {Y.}~\bibnamefont {Liu}}, \bibinfo {author} {\bibfnamefont
  {S.~M.}\ \bibnamefont {Kamali}}, \bibinfo {author} {\bibfnamefont
  {A.}~\bibnamefont {Arbabi}}, \bibinfo {author} {\bibfnamefont
  {H.}~\bibnamefont {Ruan}}, \bibinfo {author} {\bibfnamefont {A.}~\bibnamefont
  {Faraon}}, \ and\ \bibinfo {author} {\bibfnamefont {C.}~\bibnamefont
  {Yang}},\ }\href {\doibase 10.1038/s41566-017-0078-z} {\bibfield  {journal}
  {\bibinfo  {journal} {Nat. Photon.}\ }\textbf {\bibinfo {volume} {12}},\
  \bibinfo {pages} {84} (\bibinfo {year} {2018})}\BibitemShut {NoStop}%
\bibitem [{\citenamefont {McCoy}\ \emph {et~al.}(2018)\citenamefont {McCoy},
  \citenamefont {Feo}, \citenamefont {Harvey},\ and\ \citenamefont {Prum}}]{8}%
  \BibitemOpen
  \bibfield  {author} {\bibinfo {author} {\bibfnamefont {D.~E.}\ \bibnamefont
  {McCoy}}, \bibinfo {author} {\bibfnamefont {T.}~\bibnamefont {Feo}}, \bibinfo
  {author} {\bibfnamefont {T.~A.}\ \bibnamefont {Harvey}}, \ and\ \bibinfo
  {author} {\bibfnamefont {R.~O.}\ \bibnamefont {Prum}},\ }\href {\doibase
  10.1038/s41467-017-02088-w} {\bibfield  {journal} {\bibinfo  {journal} {Nat.
  Commun.}\ }\textbf {\bibinfo {volume} {9}},\ \bibinfo {pages} {1} (\bibinfo
  {year} {2018})}\BibitemShut {NoStop}%
\bibitem [{\citenamefont {Gaio}\ \emph {et~al.}(2019)\citenamefont {Gaio},
  \citenamefont {Saxena}, \citenamefont {Bertolotti}, \citenamefont
  {Pisignano}, \citenamefont {Camposeo},\ and\ \citenamefont {Sapienza}}]{9}%
  \BibitemOpen
  \bibfield  {author} {\bibinfo {author} {\bibfnamefont {M.}~\bibnamefont
  {Gaio}}, \bibinfo {author} {\bibfnamefont {D.}~\bibnamefont {Saxena}},
  \bibinfo {author} {\bibfnamefont {J.}~\bibnamefont {Bertolotti}}, \bibinfo
  {author} {\bibfnamefont {D.}~\bibnamefont {Pisignano}}, \bibinfo {author}
  {\bibfnamefont {A.}~\bibnamefont {Camposeo}}, \ and\ \bibinfo {author}
  {\bibfnamefont {R.}~\bibnamefont {Sapienza}},\ }\href {\doibase
  10.1038/s41467-018-08132-7} {\bibfield  {journal} {\bibinfo  {journal} {Nat.
  Commun.}\ }\textbf {\bibinfo {volume} {10}},\ \bibinfo {pages} {226}
  (\bibinfo {year} {2019})}\BibitemShut {NoStop}%
\bibitem [{\citenamefont {Muller}\ \emph {et~al.}(2017)\citenamefont {Muller},
  \citenamefont {Haberko}, \citenamefont {Marichy},\ and\ \citenamefont
  {Scheffold}}]{10}%
  \BibitemOpen
  \bibfield  {author} {\bibinfo {author} {\bibfnamefont {N.}~\bibnamefont
  {Muller}}, \bibinfo {author} {\bibfnamefont {J.}~\bibnamefont {Haberko}},
  \bibinfo {author} {\bibfnamefont {C.}~\bibnamefont {Marichy}}, \ and\
  \bibinfo {author} {\bibfnamefont {F.}~\bibnamefont {Scheffold}},\ }\href
  {\doibase 10.1364/OPTICA.4.000361} {\bibfield  {journal} {\bibinfo  {journal}
  {Optica}\ }\textbf {\bibinfo {volume} {4}},\ \bibinfo {pages} {361} (\bibinfo
  {year} {2017})}\BibitemShut {NoStop}%
\bibitem [{\citenamefont {Sapienza}(2019)}]{11}%
  \BibitemOpen
  \bibfield  {author} {\bibinfo {author} {\bibfnamefont {R.}~\bibnamefont
  {Sapienza}},\ }\href {\doibase 10.1038/s42254-019-0113-8} {\bibfield
  {journal} {\bibinfo  {journal} {Nat. Rev. Phys.}\ }\textbf {\bibinfo {volume}
  {1}},\ \bibinfo {pages} {690} (\bibinfo {year} {2019})}\BibitemShut {NoStop}%
\bibitem [{\citenamefont {Teimourpour}\ \emph {et~al.}(2016)\citenamefont
  {Teimourpour}, \citenamefont {Ge}, \citenamefont {Christodoulides},\ and\
  \citenamefont {El-Ganainy}}]{12}%
  \BibitemOpen
  \bibfield  {author} {\bibinfo {author} {\bibfnamefont {M.~H.}\ \bibnamefont
  {Teimourpour}}, \bibinfo {author} {\bibfnamefont {L.}~\bibnamefont {Ge}},
  \bibinfo {author} {\bibfnamefont {D.~N.}\ \bibnamefont {Christodoulides}}, \
  and\ \bibinfo {author} {\bibfnamefont {R.}~\bibnamefont {El-Ganainy}},\
  }\href {\doibase 10.1038/srep33253} {\bibfield  {journal} {\bibinfo
  {journal} {Sci. Rep.}\ }\textbf {\bibinfo {volume} {6}},\ \bibinfo {pages}
  {33253} (\bibinfo {year} {2016})}\BibitemShut {NoStop}%
\bibitem [{\citenamefont {Hughes}\ \emph {et~al.}(2005)\citenamefont {Hughes},
  \citenamefont {Ramunno}, \citenamefont {Young},\ and\ \citenamefont
  {Sipe}}]{13}%
  \BibitemOpen
  \bibfield  {author} {\bibinfo {author} {\bibfnamefont {S.}~\bibnamefont
  {Hughes}}, \bibinfo {author} {\bibfnamefont {L.}~\bibnamefont {Ramunno}},
  \bibinfo {author} {\bibfnamefont {J.~F.}\ \bibnamefont {Young}}, \ and\
  \bibinfo {author} {\bibfnamefont {J.~E.}\ \bibnamefont {Sipe}},\ }\href
  {\doibase 10.1103/PhysRevLett.94.033903} {\bibfield  {journal} {\bibinfo
  {journal} {Phys. Rev. Lett.}\ }\textbf {\bibinfo {volume} {94}},\ \bibinfo
  {pages} {033903} (\bibinfo {year} {2005})}\BibitemShut {NoStop}%
\bibitem [{\citenamefont {Jiang}\ \emph {et~al.}(2017)\citenamefont {Jiang},
  \citenamefont {Shao}, \citenamefont {Zhang}, \citenamefont {Yi},
  \citenamefont {Wiersig}, \citenamefont {Wang}, \citenamefont {Gong},
  \citenamefont {Lon{\v{c}}ar}, \citenamefont {Yang},\ and\ \citenamefont
  {Xiao}}]{14}%
  \BibitemOpen
  \bibfield  {author} {\bibinfo {author} {\bibfnamefont {X.}~\bibnamefont
  {Jiang}}, \bibinfo {author} {\bibfnamefont {L.}~\bibnamefont {Shao}},
  \bibinfo {author} {\bibfnamefont {S.-X.}\ \bibnamefont {Zhang}}, \bibinfo
  {author} {\bibfnamefont {X.}~\bibnamefont {Yi}}, \bibinfo {author}
  {\bibfnamefont {J.}~\bibnamefont {Wiersig}}, \bibinfo {author} {\bibfnamefont
  {L.}~\bibnamefont {Wang}}, \bibinfo {author} {\bibfnamefont {Q.}~\bibnamefont
  {Gong}}, \bibinfo {author} {\bibfnamefont {M.}~\bibnamefont {Lon{\v{c}}ar}},
  \bibinfo {author} {\bibfnamefont {L.}~\bibnamefont {Yang}}, \ and\ \bibinfo
  {author} {\bibfnamefont {Y.-F.}\ \bibnamefont {Xiao}},\ }\href {\doibase
  doi:10.1126/science.aao0763} {\bibfield  {journal} {\bibinfo  {journal}
  {Science}\ }\textbf {\bibinfo {volume} {358}},\ \bibinfo {pages} {344}
  (\bibinfo {year} {2017})}\BibitemShut {NoStop}%
\bibitem [{\citenamefont {Ma}\ \emph {et~al.}(2016)\citenamefont {Ma},
  \citenamefont {Guerboukha}, \citenamefont {Girard}, \citenamefont {Squires},
  \citenamefont {Lewis},\ and\ \citenamefont {Skorobogatiy}}]{15}%
  \BibitemOpen
  \bibfield  {author} {\bibinfo {author} {\bibfnamefont {T.}~\bibnamefont
  {Ma}}, \bibinfo {author} {\bibfnamefont {H.}~\bibnamefont {Guerboukha}},
  \bibinfo {author} {\bibfnamefont {M.}~\bibnamefont {Girard}}, \bibinfo
  {author} {\bibfnamefont {A.~D.}\ \bibnamefont {Squires}}, \bibinfo {author}
  {\bibfnamefont {R.~A.}\ \bibnamefont {Lewis}}, \ and\ \bibinfo {author}
  {\bibfnamefont {M.}~\bibnamefont {Skorobogatiy}},\ }\href {\doibase
  https://doi.org/10.1002/adom.201600171} {\bibfield  {journal} {\bibinfo
  {journal} {Adv. Opt. Mater.}\ }\textbf {\bibinfo {volume} {4}},\ \bibinfo
  {pages} {2085} (\bibinfo {year} {2016})}\BibitemShut {NoStop}%
\bibitem [{\citenamefont {Man}\ \emph {et~al.}(2013)\citenamefont {Man},
  \citenamefont {Florescu}, \citenamefont {Williamson}, \citenamefont {He},
  \citenamefont {Hashemizad}, \citenamefont {Leung}, \citenamefont {Liner},
  \citenamefont {Torquato}, \citenamefont {Chaikin},\ and\ \citenamefont
  {Steinhardt}}]{16}%
  \BibitemOpen
  \bibfield  {author} {\bibinfo {author} {\bibfnamefont {W.}~\bibnamefont
  {Man}}, \bibinfo {author} {\bibfnamefont {M.}~\bibnamefont {Florescu}},
  \bibinfo {author} {\bibfnamefont {E.~P.}\ \bibnamefont {Williamson}},
  \bibinfo {author} {\bibfnamefont {Y.}~\bibnamefont {He}}, \bibinfo {author}
  {\bibfnamefont {S.~R.}\ \bibnamefont {Hashemizad}}, \bibinfo {author}
  {\bibfnamefont {B.~Y.~C.}\ \bibnamefont {Leung}}, \bibinfo {author}
  {\bibfnamefont {D.~R.}\ \bibnamefont {Liner}}, \bibinfo {author}
  {\bibfnamefont {S.}~\bibnamefont {Torquato}}, \bibinfo {author}
  {\bibfnamefont {P.~M.}\ \bibnamefont {Chaikin}}, \ and\ \bibinfo {author}
  {\bibfnamefont {P.~J.}\ \bibnamefont {Steinhardt}},\ }\href {\doibase
  doi:10.1073/pnas.1307879110} {\bibfield  {journal} {\bibinfo  {journal}
  {Proc. Natl. Acad. Sci.}\ }\textbf {\bibinfo {volume} {110}},\ \bibinfo
  {pages} {15886} (\bibinfo {year} {2013})}\BibitemShut {NoStop}%
\bibitem [{\citenamefont {Cui}\ \emph {et~al.}(2022)\citenamefont {Cui},
  \citenamefont {Zhang}, \citenamefont {Zhang},\ and\ \citenamefont
  {Chan}}]{17}%
  \BibitemOpen
  \bibfield  {author} {\bibinfo {author} {\bibfnamefont {X.}~\bibnamefont
  {Cui}}, \bibinfo {author} {\bibfnamefont {R.-Y.}\ \bibnamefont {Zhang}},
  \bibinfo {author} {\bibfnamefont {Z.-Q.}\ \bibnamefont {Zhang}}, \ and\
  \bibinfo {author} {\bibfnamefont {C.~T.}\ \bibnamefont {Chan}},\ }\href
  {\doibase 10.1103/PhysRevLett.129.043902} {\bibfield  {journal} {\bibinfo
  {journal} {Phys. Rev. Lett.}\ }\textbf {\bibinfo {volume} {129}},\ \bibinfo
  {pages} {043902} (\bibinfo {year} {2022})}\BibitemShut {NoStop}%
\bibitem [{\citenamefont {Makris}\ \emph {et~al.}(2017)\citenamefont {Makris},
  \citenamefont {Brandst{\"o}tter}, \citenamefont {Ambichl}, \citenamefont
  {Musslimani},\ and\ \citenamefont {Rotter}}]{18}%
  \BibitemOpen
  \bibfield  {author} {\bibinfo {author} {\bibfnamefont {K.~G.}\ \bibnamefont
  {Makris}}, \bibinfo {author} {\bibfnamefont {A.}~\bibnamefont
  {Brandst{\"o}tter}}, \bibinfo {author} {\bibfnamefont {P.}~\bibnamefont
  {Ambichl}}, \bibinfo {author} {\bibfnamefont {Z.~H.}\ \bibnamefont
  {Musslimani}}, \ and\ \bibinfo {author} {\bibfnamefont {S.}~\bibnamefont
  {Rotter}},\ }\href {\doibase 10.1038/lsa.2017.35} {\bibfield  {journal}
  {\bibinfo  {journal} {Light Sci. Appl.}\ }\textbf {\bibinfo {volume} {6}},\
  \bibinfo {pages} {e17035} (\bibinfo {year} {2017})}\BibitemShut {NoStop}%
\bibitem [{\citenamefont {Segev}\ \emph {et~al.}(2013)\citenamefont {Segev},
  \citenamefont {Silberberg},\ and\ \citenamefont {Christodoulides}}]{19}%
  \BibitemOpen
  \bibfield  {author} {\bibinfo {author} {\bibfnamefont {M.}~\bibnamefont
  {Segev}}, \bibinfo {author} {\bibfnamefont {Y.}~\bibnamefont {Silberberg}}, \
  and\ \bibinfo {author} {\bibfnamefont {D.~N.}\ \bibnamefont
  {Christodoulides}},\ }\href {\doibase 10.1038/nphoton.2013.30} {\bibfield
  {journal} {\bibinfo  {journal} {Nat. Photon.}\ }\textbf {\bibinfo {volume}
  {7}},\ \bibinfo {pages} {197} (\bibinfo {year} {2013})}\BibitemShut {NoStop}%
\bibitem [{\citenamefont {Sharabi}\ \emph {et~al.}(2021)\citenamefont
  {Sharabi}, \citenamefont {Lustig},\ and\ \citenamefont {Segev}}]{20}%
  \BibitemOpen
  \bibfield  {author} {\bibinfo {author} {\bibfnamefont {Y.}~\bibnamefont
  {Sharabi}}, \bibinfo {author} {\bibfnamefont {E.}~\bibnamefont {Lustig}}, \
  and\ \bibinfo {author} {\bibfnamefont {M.}~\bibnamefont {Segev}},\ }\href
  {\doibase 10.1103/PhysRevLett.126.163902} {\bibfield  {journal} {\bibinfo
  {journal} {Phys. Rev. Lett.}\ }\textbf {\bibinfo {volume} {126}},\ \bibinfo
  {pages} {163902} (\bibinfo {year} {2021})}\BibitemShut {NoStop}%
\bibitem [{\citenamefont {St{\"u}tzer}\ \emph {et~al.}(2018)\citenamefont
  {St{\"u}tzer}, \citenamefont {Plotnik}, \citenamefont {Lumer}, \citenamefont
  {Titum}, \citenamefont {Lindner}, \citenamefont {Segev}, \citenamefont
  {Rechtsman},\ and\ \citenamefont {Szameit}}]{21}%
  \BibitemOpen
  \bibfield  {author} {\bibinfo {author} {\bibfnamefont {S.}~\bibnamefont
  {St{\"u}tzer}}, \bibinfo {author} {\bibfnamefont {Y.}~\bibnamefont
  {Plotnik}}, \bibinfo {author} {\bibfnamefont {Y.}~\bibnamefont {Lumer}},
  \bibinfo {author} {\bibfnamefont {P.}~\bibnamefont {Titum}}, \bibinfo
  {author} {\bibfnamefont {N.~H.}\ \bibnamefont {Lindner}}, \bibinfo {author}
  {\bibfnamefont {M.}~\bibnamefont {Segev}}, \bibinfo {author} {\bibfnamefont
  {M.~C.}\ \bibnamefont {Rechtsman}}, \ and\ \bibinfo {author} {\bibfnamefont
  {A.}~\bibnamefont {Szameit}},\ }\href {\doibase 10.1038/s41586-018-0418-2}
  {\bibfield  {journal} {\bibinfo  {journal} {Nature}\ }\textbf {\bibinfo
  {volume} {560}},\ \bibinfo {pages} {461} (\bibinfo {year}
  {2018})}\BibitemShut {NoStop}%
\bibitem [{\citenamefont {Wiersma}(2013)}]{22}%
  \BibitemOpen
  \bibfield  {author} {\bibinfo {author} {\bibfnamefont {D.~S.}\ \bibnamefont
  {Wiersma}},\ }\href {\doibase 10.1038/nphoton.2013.29} {\bibfield  {journal}
  {\bibinfo  {journal} {Nat. Photon.}\ }\textbf {\bibinfo {volume} {7}},\
  \bibinfo {pages} {188} (\bibinfo {year} {2013})}\BibitemShut {NoStop}%
\bibitem [{\citenamefont {Galiffi}\ \emph {et~al.}(2022)\citenamefont
  {Galiffi}, \citenamefont {Tirole}, \citenamefont {Yin}, \citenamefont {Li},
  \citenamefont {Vezzoli}, \citenamefont {Huidobro}, \citenamefont
  {Silveirinha}, \citenamefont {Sapienza}, \citenamefont {Al{\`u}},\ and\
  \citenamefont {Pendry}}]{23}%
  \BibitemOpen
  \bibfield  {author} {\bibinfo {author} {\bibfnamefont {E.}~\bibnamefont
  {Galiffi}}, \bibinfo {author} {\bibfnamefont {R.}~\bibnamefont {Tirole}},
  \bibinfo {author} {\bibfnamefont {S.}~\bibnamefont {Yin}}, \bibinfo {author}
  {\bibfnamefont {H.}~\bibnamefont {Li}}, \bibinfo {author} {\bibfnamefont
  {S.}~\bibnamefont {Vezzoli}}, \bibinfo {author} {\bibfnamefont {P.~A.}\
  \bibnamefont {Huidobro}}, \bibinfo {author} {\bibfnamefont {M.~G.}\
  \bibnamefont {Silveirinha}}, \bibinfo {author} {\bibfnamefont
  {R.}~\bibnamefont {Sapienza}}, \bibinfo {author} {\bibfnamefont
  {A.}~\bibnamefont {Al{\`u}}}, \ and\ \bibinfo {author} {\bibfnamefont
  {J.~B.}\ \bibnamefont {Pendry}},\ }\href {\doibase 10.1117/1.Ap.4.1.014002}
  {\bibfield  {journal} {\bibinfo  {journal} {Adv. Photonics}\ }\textbf
  {\bibinfo {volume} {4}},\ \bibinfo {pages} {014002} (\bibinfo {year}
  {2022})}\BibitemShut {NoStop}%
\bibitem [{\citenamefont {Ghatak}\ and\ \citenamefont {Das}(2019)}]{24}%
  \BibitemOpen
  \bibfield  {author} {\bibinfo {author} {\bibfnamefont {A.}~\bibnamefont
  {Ghatak}}\ and\ \bibinfo {author} {\bibfnamefont {T.}~\bibnamefont {Das}},\
  }\href {\doibase 10.1088/1361-648X/ab11b3} {\bibfield  {journal} {\bibinfo
  {journal} {J. Phys. Condens. Matter}\ }\textbf {\bibinfo {volume} {31}},\
  \bibinfo {pages} {263001} (\bibinfo {year} {2019})}\BibitemShut {NoStop}%
\bibitem [{\citenamefont {Kim}\ \emph {et~al.}(2020)\citenamefont {Kim},
  \citenamefont {Jacob},\ and\ \citenamefont {Rho}}]{25}%
  \BibitemOpen
  \bibfield  {author} {\bibinfo {author} {\bibfnamefont {M.}~\bibnamefont
  {Kim}}, \bibinfo {author} {\bibfnamefont {Z.}~\bibnamefont {Jacob}}, \ and\
  \bibinfo {author} {\bibfnamefont {J.}~\bibnamefont {Rho}},\ }\href {\doibase
  10.1038/s41377-020-0331-y} {\bibfield  {journal} {\bibinfo  {journal} {Light
  Sci. Appl.}\ }\textbf {\bibinfo {volume} {9}},\ \bibinfo {pages} {130}
  (\bibinfo {year} {2020})}\BibitemShut {NoStop}%
\bibitem [{\citenamefont {Li}\ \emph {et~al.}(2023)\citenamefont {Li},
  \citenamefont {Wei}, \citenamefont {Cotrufo}, \citenamefont {Chen},
  \citenamefont {Mann}, \citenamefont {Ni}, \citenamefont {Xu}, \citenamefont
  {Chen}, \citenamefont {Wang}, \citenamefont {Fan}, \citenamefont {Qiu},
  \citenamefont {Al{\`u}},\ and\ \citenamefont {Chen}}]{26}%
  \BibitemOpen
  \bibfield  {author} {\bibinfo {author} {\bibfnamefont {A.}~\bibnamefont
  {Li}}, \bibinfo {author} {\bibfnamefont {H.}~\bibnamefont {Wei}}, \bibinfo
  {author} {\bibfnamefont {M.}~\bibnamefont {Cotrufo}}, \bibinfo {author}
  {\bibfnamefont {W.}~\bibnamefont {Chen}}, \bibinfo {author} {\bibfnamefont
  {S.}~\bibnamefont {Mann}}, \bibinfo {author} {\bibfnamefont {X.}~\bibnamefont
  {Ni}}, \bibinfo {author} {\bibfnamefont {B.}~\bibnamefont {Xu}}, \bibinfo
  {author} {\bibfnamefont {J.}~\bibnamefont {Chen}}, \bibinfo {author}
  {\bibfnamefont {J.}~\bibnamefont {Wang}}, \bibinfo {author} {\bibfnamefont
  {S.}~\bibnamefont {Fan}}, \bibinfo {author} {\bibfnamefont {C.-W.}\
  \bibnamefont {Qiu}}, \bibinfo {author} {\bibfnamefont {A.}~\bibnamefont
  {Al{\`u}}}, \ and\ \bibinfo {author} {\bibfnamefont {L.}~\bibnamefont
  {Chen}},\ }\href {\doibase 10.1038/s41565-023-01408-0} {\bibfield  {journal}
  {\bibinfo  {journal} {Nat. Nanotechnol.}\ }\textbf {\bibinfo {volume} {18}},\
  \bibinfo {pages} {706} (\bibinfo {year} {2023})}\BibitemShut {NoStop}%
\bibitem [{\citenamefont {Lin}\ \emph {et~al.}(2023)\citenamefont {Lin},
  \citenamefont {Wang}, \citenamefont {Liu}, \citenamefont {Xue}, \citenamefont
  {Zhang}, \citenamefont {Chong},\ and\ \citenamefont {Jiang}}]{27}%
  \BibitemOpen
  \bibfield  {author} {\bibinfo {author} {\bibfnamefont {Z.-K.}\ \bibnamefont
  {Lin}}, \bibinfo {author} {\bibfnamefont {Q.}~\bibnamefont {Wang}}, \bibinfo
  {author} {\bibfnamefont {Y.}~\bibnamefont {Liu}}, \bibinfo {author}
  {\bibfnamefont {H.}~\bibnamefont {Xue}}, \bibinfo {author} {\bibfnamefont
  {B.}~\bibnamefont {Zhang}}, \bibinfo {author} {\bibfnamefont
  {Y.}~\bibnamefont {Chong}}, \ and\ \bibinfo {author} {\bibfnamefont {J.-H.}\
  \bibnamefont {Jiang}},\ }\href {\doibase 10.1038/s42254-023-00602-2}
  {\bibfield  {journal} {\bibinfo  {journal} {Nat. Rev. Phys.}\ }\textbf
  {\bibinfo {volume} {5}},\ \bibinfo {pages} {483} (\bibinfo {year}
  {2023})}\BibitemShut {NoStop}%
\bibitem [{\citenamefont {Lu}\ \emph {et~al.}(2014)\citenamefont {Lu},
  \citenamefont {Joannopoulos},\ and\ \citenamefont {Solja{\v{c}}i{\'c}}}]{28}%
  \BibitemOpen
  \bibfield  {author} {\bibinfo {author} {\bibfnamefont {L.}~\bibnamefont
  {Lu}}, \bibinfo {author} {\bibfnamefont {J.~D.}\ \bibnamefont
  {Joannopoulos}}, \ and\ \bibinfo {author} {\bibfnamefont {M.}~\bibnamefont
  {Solja{\v{c}}i{\'c}}},\ }\href {\doibase 10.1038/nphoton.2014.248} {\bibfield
   {journal} {\bibinfo  {journal} {Nat. Photon.}\ }\textbf {\bibinfo {volume}
  {8}},\ \bibinfo {pages} {821} (\bibinfo {year} {2014})}\BibitemShut {NoStop}%
\bibitem [{\citenamefont {Lustig}\ and\ \citenamefont {Segev}(2021)}]{29}%
  \BibitemOpen
  \bibfield  {author} {\bibinfo {author} {\bibfnamefont {E.}~\bibnamefont
  {Lustig}}\ and\ \bibinfo {author} {\bibfnamefont {M.}~\bibnamefont {Segev}},\
  }\href {\doibase 10.1364/AOP.418074} {\bibfield  {journal} {\bibinfo
  {journal} {Adv. Opt. Photonics}\ }\textbf {\bibinfo {volume} {13}},\ \bibinfo
  {pages} {426} (\bibinfo {year} {2021})}\BibitemShut {NoStop}%
\bibitem [{\citenamefont {Ma}\ \emph {et~al.}(2023)\citenamefont {Ma},
  \citenamefont {Zhou}, \citenamefont {Tang}, \citenamefont {Li}, \citenamefont
  {Zhang}, \citenamefont {Xi}, \citenamefont {Martin}, \citenamefont {Baron},
  \citenamefont {Liu}, \citenamefont {Zhang}, \citenamefont {Chen},\ and\
  \citenamefont {Sun}}]{30}%
  \BibitemOpen
  \bibfield  {author} {\bibinfo {author} {\bibfnamefont {J.}~\bibnamefont
  {Ma}}, \bibinfo {author} {\bibfnamefont {T.}~\bibnamefont {Zhou}}, \bibinfo
  {author} {\bibfnamefont {M.}~\bibnamefont {Tang}}, \bibinfo {author}
  {\bibfnamefont {H.}~\bibnamefont {Li}}, \bibinfo {author} {\bibfnamefont
  {Z.}~\bibnamefont {Zhang}}, \bibinfo {author} {\bibfnamefont
  {X.}~\bibnamefont {Xi}}, \bibinfo {author} {\bibfnamefont {M.}~\bibnamefont
  {Martin}}, \bibinfo {author} {\bibfnamefont {T.}~\bibnamefont {Baron}},
  \bibinfo {author} {\bibfnamefont {H.}~\bibnamefont {Liu}}, \bibinfo {author}
  {\bibfnamefont {Z.}~\bibnamefont {Zhang}}, \bibinfo {author} {\bibfnamefont
  {S.}~\bibnamefont {Chen}}, \ and\ \bibinfo {author} {\bibfnamefont
  {X.}~\bibnamefont {Sun}},\ }\href {\doibase 10.1038/s41377-023-01290-4}
  {\bibfield  {journal} {\bibinfo  {journal} {Light Sci. Appl.}\ }\textbf
  {\bibinfo {volume} {12}},\ \bibinfo {pages} {255} (\bibinfo {year}
  {2023})}\BibitemShut {NoStop}%
\bibitem [{\citenamefont {Ni}\ \emph {et~al.}(2023)\citenamefont {Ni},
  \citenamefont {Yves}, \citenamefont {Krasnok},\ and\ \citenamefont
  {Al{\`u}}}]{31}%
  \BibitemOpen
  \bibfield  {author} {\bibinfo {author} {\bibfnamefont {X.}~\bibnamefont
  {Ni}}, \bibinfo {author} {\bibfnamefont {S.}~\bibnamefont {Yves}}, \bibinfo
  {author} {\bibfnamefont {A.}~\bibnamefont {Krasnok}}, \ and\ \bibinfo
  {author} {\bibfnamefont {A.}~\bibnamefont {Al{\`u}}},\ }\href {\doibase
  10.1021/acs.chemrev.2c00800} {\bibfield  {journal} {\bibinfo  {journal}
  {Chem. Rev.}\ }\textbf {\bibinfo {volume} {123}},\ \bibinfo {pages} {7585}
  (\bibinfo {year} {2023})}\BibitemShut {NoStop}%
\bibitem [{\citenamefont {Ozawa}\ \emph {et~al.}(2019)\citenamefont {Ozawa},
  \citenamefont {Price}, \citenamefont {Amo}, \citenamefont {Goldman},
  \citenamefont {Hafezi}, \citenamefont {Lu}, \citenamefont {Rechtsman},
  \citenamefont {Schuster}, \citenamefont {Simon}, \citenamefont {Zilberberg},\
  and\ \citenamefont {Carusotto}}]{32}%
  \BibitemOpen
  \bibfield  {author} {\bibinfo {author} {\bibfnamefont {T.}~\bibnamefont
  {Ozawa}}, \bibinfo {author} {\bibfnamefont {H.~M.}\ \bibnamefont {Price}},
  \bibinfo {author} {\bibfnamefont {A.}~\bibnamefont {Amo}}, \bibinfo {author}
  {\bibfnamefont {N.}~\bibnamefont {Goldman}}, \bibinfo {author} {\bibfnamefont
  {M.}~\bibnamefont {Hafezi}}, \bibinfo {author} {\bibfnamefont
  {L.}~\bibnamefont {Lu}}, \bibinfo {author} {\bibfnamefont {M.~C.}\
  \bibnamefont {Rechtsman}}, \bibinfo {author} {\bibfnamefont {D.}~\bibnamefont
  {Schuster}}, \bibinfo {author} {\bibfnamefont {J.}~\bibnamefont {Simon}},
  \bibinfo {author} {\bibfnamefont {O.}~\bibnamefont {Zilberberg}}, \ and\
  \bibinfo {author} {\bibfnamefont {I.}~\bibnamefont {Carusotto}},\ }\href
  {\doibase 10.1103/RevModPhys.91.015006} {\bibfield  {journal} {\bibinfo
  {journal} {Rev. Mod. Phys.}\ }\textbf {\bibinfo {volume} {91}},\ \bibinfo
  {pages} {015006} (\bibinfo {year} {2019})}\BibitemShut {NoStop}%
\bibitem [{\citenamefont {Qu}\ \emph {et~al.}(2024)\citenamefont {Qu},
  \citenamefont {Wang}, \citenamefont {Cheng}, \citenamefont {Cui},
  \citenamefont {Zhang}, \citenamefont {Zhang}, \citenamefont {Zhang},
  \citenamefont {Chen},\ and\ \citenamefont {Chan}}]{33}%
  \BibitemOpen
  \bibfield  {author} {\bibinfo {author} {\bibfnamefont {T.}~\bibnamefont
  {Qu}}, \bibinfo {author} {\bibfnamefont {M.}~\bibnamefont {Wang}}, \bibinfo
  {author} {\bibfnamefont {X.}~\bibnamefont {Cheng}}, \bibinfo {author}
  {\bibfnamefont {X.}~\bibnamefont {Cui}}, \bibinfo {author} {\bibfnamefont
  {R.-Y.}\ \bibnamefont {Zhang}}, \bibinfo {author} {\bibfnamefont {Z.-Q.}\
  \bibnamefont {Zhang}}, \bibinfo {author} {\bibfnamefont {L.}~\bibnamefont
  {Zhang}}, \bibinfo {author} {\bibfnamefont {J.}~\bibnamefont {Chen}}, \ and\
  \bibinfo {author} {\bibfnamefont {C.~T.}\ \bibnamefont {Chan}},\ }\href
  {\doibase 10.1103/PhysRevLett.132.223802} {\bibfield  {journal} {\bibinfo
  {journal} {Phys. Rev. Lett.}\ }\textbf {\bibinfo {volume} {132}},\ \bibinfo
  {pages} {223802} (\bibinfo {year} {2024})}\BibitemShut {NoStop}%
\bibitem [{\citenamefont {Rudner}\ and\ \citenamefont {Lindner}(2020)}]{34}%
  \BibitemOpen
  \bibfield  {author} {\bibinfo {author} {\bibfnamefont {M.~S.}\ \bibnamefont
  {Rudner}}\ and\ \bibinfo {author} {\bibfnamefont {N.~H.}\ \bibnamefont
  {Lindner}},\ }\href {\doibase 10.1038/s42254-020-0170-z} {\bibfield
  {journal} {\bibinfo  {journal} {Nat. Rev. Phys.}\ }\textbf {\bibinfo {volume}
  {2}},\ \bibinfo {pages} {229} (\bibinfo {year} {2020})}\BibitemShut {NoStop}%
\bibitem [{\citenamefont {Segev}\ and\ \citenamefont {Bandres}(2021)}]{35}%
  \BibitemOpen
  \bibfield  {author} {\bibinfo {author} {\bibfnamefont {M.}~\bibnamefont
  {Segev}}\ and\ \bibinfo {author} {\bibfnamefont {M.~A.}\ \bibnamefont
  {Bandres}},\ }\href {\doibase doi:10.1515/nanoph-2020-0441} {\bibfield
  {journal} {\bibinfo  {journal} {Nanophotonics}\ }\textbf {\bibinfo {volume}
  {10}},\ \bibinfo {pages} {425} (\bibinfo {year} {2021})}\BibitemShut
  {NoStop}%
\bibitem [{\citenamefont {Szameit}\ and\ \citenamefont {Rechtsman}(2024)}]{36}%
  \BibitemOpen
  \bibfield  {author} {\bibinfo {author} {\bibfnamefont {A.}~\bibnamefont
  {Szameit}}\ and\ \bibinfo {author} {\bibfnamefont {M.~C.}\ \bibnamefont
  {Rechtsman}},\ }\href {\doibase 10.1038/s41567-024-02454-8} {\bibfield
  {journal} {\bibinfo  {journal} {Nat. Phys.}\ }\textbf {\bibinfo {volume}
  {20}},\ \bibinfo {pages} {905} (\bibinfo {year} {2024})}\BibitemShut
  {NoStop}%
\bibitem [{\citenamefont {Tang}\ \emph {et~al.}(2022)\citenamefont {Tang},
  \citenamefont {He}, \citenamefont {Shi}, \citenamefont {Liu}, \citenamefont
  {Chen},\ and\ \citenamefont {Dong}}]{37}%
  \BibitemOpen
  \bibfield  {author} {\bibinfo {author} {\bibfnamefont {G.-J.}\ \bibnamefont
  {Tang}}, \bibinfo {author} {\bibfnamefont {X.-T.}\ \bibnamefont {He}},
  \bibinfo {author} {\bibfnamefont {F.-L.}\ \bibnamefont {Shi}}, \bibinfo
  {author} {\bibfnamefont {J.-W.}\ \bibnamefont {Liu}}, \bibinfo {author}
  {\bibfnamefont {X.-D.}\ \bibnamefont {Chen}}, \ and\ \bibinfo {author}
  {\bibfnamefont {J.-W.}\ \bibnamefont {Dong}},\ }\href {\doibase
  https://doi.org/10.1002/lpor.202100300} {\bibfield  {journal} {\bibinfo
  {journal} {Laser Photonics Rev.}\ }\textbf {\bibinfo {volume} {16}},\
  \bibinfo {pages} {2100300} (\bibinfo {year} {2022})}\BibitemShut {NoStop}%
\bibitem [{\citenamefont {Zhao}\ \emph {et~al.}(2022)\citenamefont {Zhao},
  \citenamefont {Jin}, \citenamefont {Liu}, \citenamefont {Chao}, \citenamefont
  {Liu}, \citenamefont {Zhang}, \citenamefont {Wang}, \citenamefont {Lyu},
  \citenamefont {Wageh}, \citenamefont {Al-Hartomy}, \citenamefont {Al-Sehemi},
  \citenamefont {Fu},\ and\ \citenamefont {Zhang}}]{38}%
  \BibitemOpen
  \bibfield  {author} {\bibinfo {author} {\bibfnamefont {X.}~\bibnamefont
  {Zhao}}, \bibinfo {author} {\bibfnamefont {H.}~\bibnamefont {Jin}}, \bibinfo
  {author} {\bibfnamefont {J.}~\bibnamefont {Liu}}, \bibinfo {author}
  {\bibfnamefont {J.}~\bibnamefont {Chao}}, \bibinfo {author} {\bibfnamefont
  {T.}~\bibnamefont {Liu}}, \bibinfo {author} {\bibfnamefont {H.}~\bibnamefont
  {Zhang}}, \bibinfo {author} {\bibfnamefont {G.}~\bibnamefont {Wang}},
  \bibinfo {author} {\bibfnamefont {W.}~\bibnamefont {Lyu}}, \bibinfo {author}
  {\bibfnamefont {S.}~\bibnamefont {Wageh}}, \bibinfo {author} {\bibfnamefont
  {O.~A.}\ \bibnamefont {Al-Hartomy}}, \bibinfo {author} {\bibfnamefont
  {A.~G.}\ \bibnamefont {Al-Sehemi}}, \bibinfo {author} {\bibfnamefont
  {B.}~\bibnamefont {Fu}}, \ and\ \bibinfo {author} {\bibfnamefont
  {H.}~\bibnamefont {Zhang}},\ }\href {\doibase
  https://doi.org/10.1002/lpor.202200386} {\bibfield  {journal} {\bibinfo
  {journal} {Laser Photonics Rev.}\ }\textbf {\bibinfo {volume} {16}},\
  \bibinfo {pages} {2200386} (\bibinfo {year} {2022})}\BibitemShut {NoStop}%
\bibitem [{\citenamefont {Chen}\ \emph {et~al.}(2024)\citenamefont {Chen},
  \citenamefont {Gao}, \citenamefont {Cui}, \citenamefont {Mo}, \citenamefont
  {Chen}, \citenamefont {Zhang}, \citenamefont {Chan},\ and\ \citenamefont
  {Dong}}]{39}%
  \BibitemOpen
  \bibfield  {author} {\bibinfo {author} {\bibfnamefont {X.-D.}\ \bibnamefont
  {Chen}}, \bibinfo {author} {\bibfnamefont {Z.-X.}\ \bibnamefont {Gao}},
  \bibinfo {author} {\bibfnamefont {X.}~\bibnamefont {Cui}}, \bibinfo {author}
  {\bibfnamefont {H.-C.}\ \bibnamefont {Mo}}, \bibinfo {author} {\bibfnamefont
  {W.-J.}\ \bibnamefont {Chen}}, \bibinfo {author} {\bibfnamefont {R.-Y.}\
  \bibnamefont {Zhang}}, \bibinfo {author} {\bibfnamefont {C.~T.}\ \bibnamefont
  {Chan}}, \ and\ \bibinfo {author} {\bibfnamefont {J.-W.}\ \bibnamefont
  {Dong}},\ }\href {\doibase 10.1103/PhysRevLett.133.133802} {\bibfield
  {journal} {\bibinfo  {journal} {Phys. Rev. Lett.}\ }\textbf {\bibinfo
  {volume} {133}},\ \bibinfo {pages} {133802} (\bibinfo {year}
  {2024})}\BibitemShut {NoStop}%
\bibitem [{\citenamefont {Cheng}\ \emph {et~al.}(2022)\citenamefont {Cheng},
  \citenamefont {Liu}, \citenamefont {Liu},\ and\ \citenamefont {Chen}}]{40}%
  \BibitemOpen
  \bibfield  {author} {\bibinfo {author} {\bibfnamefont {W.}~\bibnamefont
  {Cheng}}, \bibinfo {author} {\bibfnamefont {W.}~\bibnamefont {Liu}}, \bibinfo
  {author} {\bibfnamefont {Q.}~\bibnamefont {Liu}}, \ and\ \bibinfo {author}
  {\bibfnamefont {F.}~\bibnamefont {Chen}},\ }\href {\doibase
  10.1364/OL.461485} {\bibfield  {journal} {\bibinfo  {journal} {Opt. Lett.}\
  }\textbf {\bibinfo {volume} {47}},\ \bibinfo {pages} {2883} (\bibinfo {year}
  {2022})}\BibitemShut {NoStop}%
\bibitem [{\citenamefont {Fang}\ \emph {et~al.}(2017)\citenamefont {Fang},
  \citenamefont {Zhang}, \citenamefont {Louie},\ and\ \citenamefont
  {Chan}}]{41}%
  \BibitemOpen
  \bibfield  {author} {\bibinfo {author} {\bibfnamefont {A.}~\bibnamefont
  {Fang}}, \bibinfo {author} {\bibfnamefont {Z.~Q.}\ \bibnamefont {Zhang}},
  \bibinfo {author} {\bibfnamefont {S.~G.}\ \bibnamefont {Louie}}, \ and\
  \bibinfo {author} {\bibfnamefont {C.~T.}\ \bibnamefont {Chan}},\ }\href
  {\doibase doi:10.1073/pnas.1620313114} {\bibfield  {journal} {\bibinfo
  {journal} {Proc. Natl. Acad. Sci. U.S.A.}\ }\textbf {\bibinfo {volume}
  {114}},\ \bibinfo {pages} {4087} (\bibinfo {year} {2017})}\BibitemShut
  {NoStop}%
\bibitem [{\citenamefont {Liu}\ \emph {et~al.}(2017)\citenamefont {Liu},
  \citenamefont {Gao}, \citenamefont {Yang},\ and\ \citenamefont {Zhang}}]{42}%
  \BibitemOpen
  \bibfield  {author} {\bibinfo {author} {\bibfnamefont {C.}~\bibnamefont
  {Liu}}, \bibinfo {author} {\bibfnamefont {W.}~\bibnamefont {Gao}}, \bibinfo
  {author} {\bibfnamefont {B.}~\bibnamefont {Yang}}, \ and\ \bibinfo {author}
  {\bibfnamefont {S.}~\bibnamefont {Zhang}},\ }\href {\doibase
  10.1103/PhysRevLett.119.183901} {\bibfield  {journal} {\bibinfo  {journal}
  {Phys. Rev. Lett.}\ }\textbf {\bibinfo {volume} {119}},\ \bibinfo {pages}
  {183901} (\bibinfo {year} {2017})}\BibitemShut {NoStop}%
\bibitem [{\citenamefont {Liu}\ \emph {et~al.}(2022{\natexlab{a}})\citenamefont
  {Liu}, \citenamefont {Zhang}, \citenamefont {Tang},\ and\ \citenamefont
  {Zhang}}]{43}%
  \BibitemOpen
  \bibfield  {author} {\bibinfo {author} {\bibfnamefont {S.-N.}\ \bibnamefont
  {Liu}}, \bibinfo {author} {\bibfnamefont {G.-Q.}\ \bibnamefont {Zhang}},
  \bibinfo {author} {\bibfnamefont {L.-Z.}\ \bibnamefont {Tang}}, \ and\
  \bibinfo {author} {\bibfnamefont {D.-W.}\ \bibnamefont {Zhang}},\ }\href
  {\doibase https://doi.org/10.1016/j.physleta.2022.128004} {\bibfield
  {journal} {\bibinfo  {journal} {Phys. Lett. A}\ }\textbf {\bibinfo {volume}
  {431}},\ \bibinfo {pages} {128004} (\bibinfo {year}
  {2022}{\natexlab{a}})}\BibitemShut {NoStop}%
\bibitem [{\citenamefont {Ren}\ \emph {et~al.}(2024)\citenamefont {Ren},
  \citenamefont {Yu}, \citenamefont {Wu}, \citenamefont {Qi}, \citenamefont
  {Wang}, \citenamefont {Yao}, \citenamefont {Ren}, \citenamefont {Guo},
  \citenamefont {Jiang}, \citenamefont {Chen}, \citenamefont {Liu},
  \citenamefont {Chen},\ and\ \citenamefont {Sun}}]{44}%
  \BibitemOpen
  \bibfield  {author} {\bibinfo {author} {\bibfnamefont {M.}~\bibnamefont
  {Ren}}, \bibinfo {author} {\bibfnamefont {Y.}~\bibnamefont {Yu}}, \bibinfo
  {author} {\bibfnamefont {B.}~\bibnamefont {Wu}}, \bibinfo {author}
  {\bibfnamefont {X.}~\bibnamefont {Qi}}, \bibinfo {author} {\bibfnamefont
  {Y.}~\bibnamefont {Wang}}, \bibinfo {author} {\bibfnamefont {X.}~\bibnamefont
  {Yao}}, \bibinfo {author} {\bibfnamefont {J.}~\bibnamefont {Ren}}, \bibinfo
  {author} {\bibfnamefont {Z.}~\bibnamefont {Guo}}, \bibinfo {author}
  {\bibfnamefont {H.}~\bibnamefont {Jiang}}, \bibinfo {author} {\bibfnamefont
  {H.}~\bibnamefont {Chen}}, \bibinfo {author} {\bibfnamefont {X.-J.}\
  \bibnamefont {Liu}}, \bibinfo {author} {\bibfnamefont {Z.}~\bibnamefont
  {Chen}}, \ and\ \bibinfo {author} {\bibfnamefont {Y.}~\bibnamefont {Sun}},\
  }\href {\doibase 10.1103/PhysRevLett.132.066602} {\bibfield  {journal}
  {\bibinfo  {journal} {Phys. Rev. Lett.}\ }\textbf {\bibinfo {volume} {132}},\
  \bibinfo {pages} {066602} (\bibinfo {year} {2024})}\BibitemShut {NoStop}%
\bibitem [{\citenamefont {Skipetrov}\ and\ \citenamefont {Wulles}(2022)}]{45}%
  \BibitemOpen
  \bibfield  {author} {\bibinfo {author} {\bibfnamefont {S.~E.}\ \bibnamefont
  {Skipetrov}}\ and\ \bibinfo {author} {\bibfnamefont {P.}~\bibnamefont
  {Wulles}},\ }\href {\doibase 10.1103/PhysRevA.105.043514} {\bibfield
  {journal} {\bibinfo  {journal} {Phys. Rev. A}\ }\textbf {\bibinfo {volume}
  {105}},\ \bibinfo {pages} {043514} (\bibinfo {year} {2022})}\BibitemShut
  {NoStop}%
\bibitem [{\citenamefont {Zhang}\ \emph {et~al.}(2022)\citenamefont {Zhang},
  \citenamefont {Sui}, \citenamefont {Zhang}, \citenamefont {Liu},
  \citenamefont {Shi}, \citenamefont {Lv}, \citenamefont {Zhang}, \citenamefont
  {Rong},\ and\ \citenamefont {Yang}}]{46}%
  \BibitemOpen
  \bibfield  {author} {\bibinfo {author} {\bibfnamefont {H.}~\bibnamefont
  {Zhang}}, \bibinfo {author} {\bibfnamefont {W.}~\bibnamefont {Sui}}, \bibinfo
  {author} {\bibfnamefont {Y.}~\bibnamefont {Zhang}}, \bibinfo {author}
  {\bibfnamefont {G.}~\bibnamefont {Liu}}, \bibinfo {author} {\bibfnamefont
  {Q.}~\bibnamefont {Shi}}, \bibinfo {author} {\bibfnamefont {Z.}~\bibnamefont
  {Lv}}, \bibinfo {author} {\bibfnamefont {D.}~\bibnamefont {Zhang}}, \bibinfo
  {author} {\bibfnamefont {C.}~\bibnamefont {Rong}}, \ and\ \bibinfo {author}
  {\bibfnamefont {B.}~\bibnamefont {Yang}},\ }\href {\doibase
  https://doi.org/10.1002/pssb.202200214} {\bibfield  {journal} {\bibinfo
  {journal} {Phys. Status Solidi B}\ }\textbf {\bibinfo {volume} {259}},\
  \bibinfo {pages} {2200214} (\bibinfo {year} {2022})}\BibitemShut {NoStop}%
\bibitem [{\citenamefont {Cheng}\ \emph {et~al.}(2020)\citenamefont {Cheng},
  \citenamefont {Wang}, \citenamefont {Niu},\ and\ \citenamefont {Zhao}}]{47}%
  \BibitemOpen
  \bibfield  {author} {\bibinfo {author} {\bibfnamefont {Y.}~\bibnamefont
  {Cheng}}, \bibinfo {author} {\bibfnamefont {Y.}~\bibnamefont {Wang}},
  \bibinfo {author} {\bibfnamefont {Y.}~\bibnamefont {Niu}}, \ and\ \bibinfo
  {author} {\bibfnamefont {Z.}~\bibnamefont {Zhao}},\ }\href {\doibase
  10.1364/OE.384029} {\bibfield  {journal} {\bibinfo  {journal} {Opt. Express}\
  }\textbf {\bibinfo {volume} {28}},\ \bibinfo {pages} {6350} (\bibinfo {year}
  {2020})}\BibitemShut {NoStop}%
\bibitem [{\citenamefont {Hu}\ \emph {et~al.}(2023)\citenamefont {Hu},
  \citenamefont {Zhan}, \citenamefont {Shi}, \citenamefont {Qiao},
  \citenamefont {He},\ and\ \citenamefont {Liu}}]{48}%
  \BibitemOpen
  \bibfield  {author} {\bibinfo {author} {\bibfnamefont {J.}~\bibnamefont
  {Hu}}, \bibinfo {author} {\bibfnamefont {C.}~\bibnamefont {Zhan}}, \bibinfo
  {author} {\bibfnamefont {H.}~\bibnamefont {Shi}}, \bibinfo {author}
  {\bibfnamefont {P.}~\bibnamefont {Qiao}}, \bibinfo {author} {\bibfnamefont
  {Y.}~\bibnamefont {He}}, \ and\ \bibinfo {author} {\bibfnamefont
  {Y.}~\bibnamefont {Liu}},\ }\href {\doibase
  https://doi.org/10.1016/j.infrared.2022.104448} {\bibfield  {journal}
  {\bibinfo  {journal} {Infrared Phys. Techn.}\ }\textbf {\bibinfo {volume}
  {131}},\ \bibinfo {pages} {104448} (\bibinfo {year} {2023})}\BibitemShut
  {NoStop}%
\bibitem [{\citenamefont {Knyazkova}\ \emph {et~al.}(2020)\citenamefont
  {Knyazkova}, \citenamefont {Borisov}, \citenamefont {Spirina},\ and\
  \citenamefont {Kistenev}}]{49}%
  \BibitemOpen
  \bibfield  {author} {\bibinfo {author} {\bibfnamefont {A.~I.}\ \bibnamefont
  {Knyazkova}}, \bibinfo {author} {\bibfnamefont {A.~V.}\ \bibnamefont
  {Borisov}}, \bibinfo {author} {\bibfnamefont {L.~V.}\ \bibnamefont
  {Spirina}}, \ and\ \bibinfo {author} {\bibfnamefont {Y.~V.}\ \bibnamefont
  {Kistenev}},\ }\href {\doibase 10.1007/s10762-020-00673-7} {\bibfield
  {journal} {\bibinfo  {journal} {J. Infrared Milli. Terahz. Waves}\ }\textbf
  {\bibinfo {volume} {41}},\ \bibinfo {pages} {1089} (\bibinfo {year}
  {2020})}\BibitemShut {NoStop}%
\bibitem [{\citenamefont {Liu}\ \emph {et~al.}(2022{\natexlab{b}})\citenamefont
  {Liu}, \citenamefont {Vohra}, \citenamefont {Bailey}, \citenamefont
  {El-Shenawee},\ and\ \citenamefont {Nelson}}]{50}%
  \BibitemOpen
  \bibfield  {author} {\bibinfo {author} {\bibfnamefont {H.}~\bibnamefont
  {Liu}}, \bibinfo {author} {\bibfnamefont {N.}~\bibnamefont {Vohra}}, \bibinfo
  {author} {\bibfnamefont {K.}~\bibnamefont {Bailey}}, \bibinfo {author}
  {\bibfnamefont {M.}~\bibnamefont {El-Shenawee}}, \ and\ \bibinfo {author}
  {\bibfnamefont {A.~H.}\ \bibnamefont {Nelson}},\ }\href {\doibase
  10.1007/s10762-021-00839-x} {\bibfield  {journal} {\bibinfo  {journal} {J.
  Infrared Milli. Terahz. Waves}\ }\textbf {\bibinfo {volume} {43}},\ \bibinfo
  {pages} {48} (\bibinfo {year} {2022}{\natexlab{b}})}\BibitemShut {NoStop}%
\bibitem [{\citenamefont {Liu}\ \emph {et~al.}(2007)\citenamefont {Liu},
  \citenamefont {Zhong}, \citenamefont {Karpowicz}, \citenamefont {Chen},\ and\
  \citenamefont {Zhang}}]{51}%
  \BibitemOpen
  \bibfield  {author} {\bibinfo {author} {\bibfnamefont {H.~B.}\ \bibnamefont
  {Liu}}, \bibinfo {author} {\bibfnamefont {H.}~\bibnamefont {Zhong}}, \bibinfo
  {author} {\bibfnamefont {N.}~\bibnamefont {Karpowicz}}, \bibinfo {author}
  {\bibfnamefont {Y.}~\bibnamefont {Chen}}, \ and\ \bibinfo {author}
  {\bibfnamefont {X.~C.}\ \bibnamefont {Zhang}},\ }\href {\doibase
  10.1109/JPROC.2007.898903} {\bibfield  {journal} {\bibinfo  {journal} {Proc.
  IEEE}\ }\textbf {\bibinfo {volume} {95}},\ \bibinfo {pages} {1514} (\bibinfo
  {year} {2007})}\BibitemShut {NoStop}%
\bibitem [{\citenamefont {Sun}\ \emph {et~al.}(2021)\citenamefont {Sun},
  \citenamefont {Li}, \citenamefont {Shen},\ and\ \citenamefont {Li}}]{52}%
  \BibitemOpen
  \bibfield  {author} {\bibinfo {author} {\bibfnamefont {X.}~\bibnamefont
  {Sun}}, \bibinfo {author} {\bibfnamefont {J.}~\bibnamefont {Li}}, \bibinfo
  {author} {\bibfnamefont {Y.}~\bibnamefont {Shen}}, \ and\ \bibinfo {author}
  {\bibfnamefont {W.}~\bibnamefont {Li}},\ }\href {\doibase
  10.3389/fnut.2021.757491} {\bibfield  {journal} {\bibinfo  {journal} {Front.
  Nutr.}\ }\textbf {\bibinfo {volume} {8}} (\bibinfo {year} {2021}),\
  10.3389/fnut.2021.757491}\BibitemShut {NoStop}%
\bibitem [{\citenamefont {Takida}\ \emph {et~al.}(2021)\citenamefont {Takida},
  \citenamefont {Nawata},\ and\ \citenamefont {Minamide}}]{53}%
  \BibitemOpen
  \bibfield  {author} {\bibinfo {author} {\bibfnamefont {Y.}~\bibnamefont
  {Takida}}, \bibinfo {author} {\bibfnamefont {K.}~\bibnamefont {Nawata}}, \
  and\ \bibinfo {author} {\bibfnamefont {H.}~\bibnamefont {Minamide}},\ }\href
  {\doibase 10.1364/oe.413201} {\bibfield  {journal} {\bibinfo  {journal} {Opt.
  Express}\ }\textbf {\bibinfo {volume} {29}},\ \bibinfo {pages} {2529}
  (\bibinfo {year} {2021})}\BibitemShut {NoStop}%
\bibitem [{\citenamefont {Tan}\ \emph {et~al.}(2022)\citenamefont {Tan},
  \citenamefont {Wang}, \citenamefont {Kumar},\ and\ \citenamefont
  {Singh}}]{54}%
  \BibitemOpen
  \bibfield  {author} {\bibinfo {author} {\bibfnamefont {Y.~J.}\ \bibnamefont
  {Tan}}, \bibinfo {author} {\bibfnamefont {W.}~\bibnamefont {Wang}}, \bibinfo
  {author} {\bibfnamefont {A.}~\bibnamefont {Kumar}}, \ and\ \bibinfo {author}
  {\bibfnamefont {R.}~\bibnamefont {Singh}},\ }\href {\doibase
  10.1364/OE.468010} {\bibfield  {journal} {\bibinfo  {journal} {Opt. Express}\
  }\textbf {\bibinfo {volume} {30}},\ \bibinfo {pages} {33035} (\bibinfo {year}
  {2022})}\BibitemShut {NoStop}%
\bibitem [{\citenamefont {Tielrooij}\ \emph {et~al.}(2022)\citenamefont
  {Tielrooij}, \citenamefont {Principi}, \citenamefont {Reig}, \citenamefont
  {Block}, \citenamefont {Varghese}, \citenamefont {Schreyeck}, \citenamefont
  {Brunner}, \citenamefont {Karczewski}, \citenamefont {Ilyakov}, \citenamefont
  {Ponomaryov}, \citenamefont {de~Oliveira}, \citenamefont {Chen},
  \citenamefont {Deinert}, \citenamefont {Carbonell}, \citenamefont
  {Valenzuela}, \citenamefont {Molenkamp}, \citenamefont {Kiessling},
  \citenamefont {Astakhov},\ and\ \citenamefont {Kovalev}}]{55}%
  \BibitemOpen
  \bibfield  {author} {\bibinfo {author} {\bibfnamefont {K.-J.}\ \bibnamefont
  {Tielrooij}}, \bibinfo {author} {\bibfnamefont {A.}~\bibnamefont {Principi}},
  \bibinfo {author} {\bibfnamefont {D.~S.}\ \bibnamefont {Reig}}, \bibinfo
  {author} {\bibfnamefont {A.}~\bibnamefont {Block}}, \bibinfo {author}
  {\bibfnamefont {S.}~\bibnamefont {Varghese}}, \bibinfo {author}
  {\bibfnamefont {S.}~\bibnamefont {Schreyeck}}, \bibinfo {author}
  {\bibfnamefont {K.}~\bibnamefont {Brunner}}, \bibinfo {author} {\bibfnamefont
  {G.}~\bibnamefont {Karczewski}}, \bibinfo {author} {\bibfnamefont
  {I.}~\bibnamefont {Ilyakov}}, \bibinfo {author} {\bibfnamefont
  {O.}~\bibnamefont {Ponomaryov}}, \bibinfo {author} {\bibfnamefont {T.~V.
  A.~G.}\ \bibnamefont {de~Oliveira}}, \bibinfo {author} {\bibfnamefont
  {M.}~\bibnamefont {Chen}}, \bibinfo {author} {\bibfnamefont {J.-C.}\
  \bibnamefont {Deinert}}, \bibinfo {author} {\bibfnamefont {C.~G.}\
  \bibnamefont {Carbonell}}, \bibinfo {author} {\bibfnamefont {S.~O.}\
  \bibnamefont {Valenzuela}}, \bibinfo {author} {\bibfnamefont {L.~W.}\
  \bibnamefont {Molenkamp}}, \bibinfo {author} {\bibfnamefont {T.}~\bibnamefont
  {Kiessling}}, \bibinfo {author} {\bibfnamefont {G.~V.}\ \bibnamefont
  {Astakhov}}, \ and\ \bibinfo {author} {\bibfnamefont {S.}~\bibnamefont
  {Kovalev}},\ }\href {\doibase 10.1038/s41377-022-01008-y} {\bibfield
  {journal} {\bibinfo  {journal} {Light Sci. Appl.}\ }\textbf {\bibinfo
  {volume} {11}},\ \bibinfo {pages} {315} (\bibinfo {year} {2022})}\BibitemShut
  {NoStop}%
\bibitem [{\citenamefont {Wan}\ \emph {et~al.}(2020)\citenamefont {Wan},
  \citenamefont {Healy},\ and\ \citenamefont {Sheridan}}]{56}%
  \BibitemOpen
  \bibfield  {author} {\bibinfo {author} {\bibfnamefont {M.}~\bibnamefont
  {Wan}}, \bibinfo {author} {\bibfnamefont {J.~J.}\ \bibnamefont {Healy}}, \
  and\ \bibinfo {author} {\bibfnamefont {J.~T.}\ \bibnamefont {Sheridan}},\
  }\href {\doibase https://doi.org/10.1016/j.optlastec.2019.105859} {\bibfield
  {journal} {\bibinfo  {journal} {Opt. Laser Technol.}\ }\textbf {\bibinfo
  {volume} {122}},\ \bibinfo {pages} {105859} (\bibinfo {year}
  {2020})}\BibitemShut {NoStop}%
\bibitem [{\citenamefont {Aberra~Guebrou}\ \emph {et~al.}(2012)\citenamefont
  {Aberra~Guebrou}, \citenamefont {Symonds}, \citenamefont {Homeyer},
  \citenamefont {Plenet}, \citenamefont {Gartstein}, \citenamefont
  {Agranovich},\ and\ \citenamefont {Bellessa}}]{57}%
  \BibitemOpen
  \bibfield  {author} {\bibinfo {author} {\bibfnamefont {S.}~\bibnamefont
  {Aberra~Guebrou}}, \bibinfo {author} {\bibfnamefont {C.}~\bibnamefont
  {Symonds}}, \bibinfo {author} {\bibfnamefont {E.}~\bibnamefont {Homeyer}},
  \bibinfo {author} {\bibfnamefont {J.~C.}\ \bibnamefont {Plenet}}, \bibinfo
  {author} {\bibfnamefont {Y.~N.}\ \bibnamefont {Gartstein}}, \bibinfo {author}
  {\bibfnamefont {V.~M.}\ \bibnamefont {Agranovich}}, \ and\ \bibinfo {author}
  {\bibfnamefont {J.}~\bibnamefont {Bellessa}},\ }\href {\doibase
  10.1103/PhysRevLett.108.066401} {\bibfield  {journal} {\bibinfo  {journal}
  {Phys. Rev. Lett.}\ }\textbf {\bibinfo {volume} {108}},\ \bibinfo {pages}
  {066401} (\bibinfo {year} {2012})}\BibitemShut {NoStop}%
\bibitem [{\citenamefont {Balasubrahmaniyam}\ \emph {et~al.}(2018)\citenamefont
  {Balasubrahmaniyam}, \citenamefont {Nahata},\ and\ \citenamefont
  {Mujumdar}}]{58}%
  \BibitemOpen
  \bibfield  {author} {\bibinfo {author} {\bibfnamefont {M.}~\bibnamefont
  {Balasubrahmaniyam}}, \bibinfo {author} {\bibfnamefont {A.}~\bibnamefont
  {Nahata}}, \ and\ \bibinfo {author} {\bibfnamefont {S.}~\bibnamefont
  {Mujumdar}},\ }\href {\doibase 10.1103/PhysRevB.98.024202} {\bibfield
  {journal} {\bibinfo  {journal} {Phys. Rev. B}\ }\textbf {\bibinfo {volume}
  {98}},\ \bibinfo {pages} {024202} (\bibinfo {year} {2018})}\BibitemShut
  {NoStop}%
\bibitem [{\citenamefont {Bozhevolnyi}\ \emph {et~al.}(2002)\citenamefont
  {Bozhevolnyi}, \citenamefont {Volkov},\ and\ \citenamefont {Leosson}}]{59}%
  \BibitemOpen
  \bibfield  {author} {\bibinfo {author} {\bibfnamefont {S.~I.}\ \bibnamefont
  {Bozhevolnyi}}, \bibinfo {author} {\bibfnamefont {V.~S.}\ \bibnamefont
  {Volkov}}, \ and\ \bibinfo {author} {\bibfnamefont {K.}~\bibnamefont
  {Leosson}},\ }\href {\doibase 10.1103/PhysRevLett.89.186801} {\bibfield
  {journal} {\bibinfo  {journal} {Phys. Rev. Lett.}\ }\textbf {\bibinfo
  {volume} {89}},\ \bibinfo {pages} {186801} (\bibinfo {year}
  {2002})}\BibitemShut {NoStop}%
\bibitem [{\citenamefont {Duan}\ \emph {et~al.}(2019)\citenamefont {Duan},
  \citenamefont {Xiao},\ and\ \citenamefont {Chen}}]{60}%
  \BibitemOpen
  \bibfield  {author} {\bibinfo {author} {\bibfnamefont {J.}~\bibnamefont
  {Duan}}, \bibinfo {author} {\bibfnamefont {S.}~\bibnamefont {Xiao}}, \ and\
  \bibinfo {author} {\bibfnamefont {J.}~\bibnamefont {Chen}},\ }\href {\doibase
  https://doi.org/10.1002/advs.201801974} {\bibfield  {journal} {\bibinfo
  {journal} {Adv. Sci.}\ }\textbf {\bibinfo {volume} {6}},\ \bibinfo {pages}
  {1801974} (\bibinfo {year} {2019})}\BibitemShut {NoStop}%
\bibitem [{\citenamefont {Herath}\ and\ \citenamefont {Premaratne}(2022)}]{61}%
  \BibitemOpen
  \bibfield  {author} {\bibinfo {author} {\bibfnamefont {K.}~\bibnamefont
  {Herath}}\ and\ \bibinfo {author} {\bibfnamefont {M.}~\bibnamefont
  {Premaratne}},\ }\href {\doibase 10.1103/PhysRevB.106.235422} {\bibfield
  {journal} {\bibinfo  {journal} {Phys. Rev. B}\ }\textbf {\bibinfo {volume}
  {106}},\ \bibinfo {pages} {235422} (\bibinfo {year} {2022})}\BibitemShut
  {NoStop}%
\bibitem [{\citenamefont {Lee}\ \emph {et~al.}(2022)\citenamefont {Lee},
  \citenamefont {Berk}, \citenamefont {Webster}, \citenamefont {Kim},\ and\
  \citenamefont {Foreman}}]{62}%
  \BibitemOpen
  \bibfield  {author} {\bibinfo {author} {\bibfnamefont {H.}~\bibnamefont
  {Lee}}, \bibinfo {author} {\bibfnamefont {J.}~\bibnamefont {Berk}}, \bibinfo
  {author} {\bibfnamefont {A.}~\bibnamefont {Webster}}, \bibinfo {author}
  {\bibfnamefont {D.}~\bibnamefont {Kim}}, \ and\ \bibinfo {author}
  {\bibfnamefont {M.~R.}\ \bibnamefont {Foreman}},\ }\href {\doibase
  10.1088/1361-6528/ac43e9} {\bibfield  {journal} {\bibinfo  {journal}
  {Nanotechnology}\ }\textbf {\bibinfo {volume} {33}},\ \bibinfo {pages}
  {165502} (\bibinfo {year} {2022})}\BibitemShut {NoStop}%
\bibitem [{\citenamefont {Petr{\'a}{\v{c}}ek}\ and\ \citenamefont
  {Kuzmiak}(2018)}]{63}%
  \BibitemOpen
  \bibfield  {author} {\bibinfo {author} {\bibfnamefont {J.}~\bibnamefont
  {Petr{\'a}{\v{c}}ek}}\ and\ \bibinfo {author} {\bibfnamefont
  {V.}~\bibnamefont {Kuzmiak}},\ }\href {\doibase 10.1103/PhysRevA.98.023806}
  {\bibfield  {journal} {\bibinfo  {journal} {Phys. Rev. A}\ }\textbf {\bibinfo
  {volume} {98}},\ \bibinfo {pages} {023806} (\bibinfo {year}
  {2018})}\BibitemShut {NoStop}%
\bibitem [{\citenamefont {Shi}\ \emph {et~al.}(2018)\citenamefont {Shi},
  \citenamefont {Liu}, \citenamefont {Peng}, \citenamefont {Xu}, \citenamefont
  {Zhang}, \citenamefont {Jing}, \citenamefont {Fan}, \citenamefont {Huang},
  \citenamefont {Wang},\ and\ \citenamefont {Wang}}]{64}%
  \BibitemOpen
  \bibfield  {author} {\bibinfo {author} {\bibfnamefont {W.-B.}\ \bibnamefont
  {Shi}}, \bibinfo {author} {\bibfnamefont {L.-Z.}\ \bibnamefont {Liu}},
  \bibinfo {author} {\bibfnamefont {R.}~\bibnamefont {Peng}}, \bibinfo {author}
  {\bibfnamefont {D.-H.}\ \bibnamefont {Xu}}, \bibinfo {author} {\bibfnamefont
  {K.}~\bibnamefont {Zhang}}, \bibinfo {author} {\bibfnamefont
  {H.}~\bibnamefont {Jing}}, \bibinfo {author} {\bibfnamefont {R.-H.}\
  \bibnamefont {Fan}}, \bibinfo {author} {\bibfnamefont {X.-R.}\ \bibnamefont
  {Huang}}, \bibinfo {author} {\bibfnamefont {Q.-J.}\ \bibnamefont {Wang}}, \
  and\ \bibinfo {author} {\bibfnamefont {M.}~\bibnamefont {Wang}},\ }\href
  {\doibase 10.1021/acs.nanolett.7b05191} {\bibfield  {journal} {\bibinfo
  {journal} {Nano Lett.}\ }\textbf {\bibinfo {volume} {18}},\ \bibinfo {pages}
  {1896} (\bibinfo {year} {2018})}\BibitemShut {NoStop}%
\end{thebibliography}%

\end{document}